\documentclass[prx,floatfix,twocolumn,showpacs]{revtex4-1}
\usepackage{amsmath}
\usepackage{graphicx}
\usepackage{amsfonts}
\usepackage{amssymb}
\usepackage{bm}
\usepackage{epsfig,float,afterpage}
\usepackage[colorlinks,linkcolor=blue,citecolor=blue,urlcolor=blue]{hyperref}
\usepackage[usenames,dvipsnames]{color}

\newcommand{\beq}{\begin{equation}}
\newcommand{\eeq}{\end{equation}}
\newcommand{\bes}{\begin{subequations}}
\newcommand{\ees}{\end{subequations}}
\newcommand{\bea}{\begin{eqnarray}}
\newcommand{\eea}{\end{eqnarray}}
\newcommand{\ben}{\begin{enumerate}}
\newcommand{\een}{\end{enumerate}}
\newcommand{\ba}{\begin{array}}
\newcommand{\ea}{\end{array}}
\newcommand{\beqn}{\begin{eqnarray*}}
\newcommand{\eeqn}{\end{eqnarray*}}
\newcommand{\bbm}{\begin{bmatrix}}
\newcommand{\ebm}{\end{bmatrix}}

\newcommand{\f}[2]{\frac{#1}{#2}}
\newcommand{\g}{\gamma}
\newcommand{\G}{\Gamma}
\newcommand{\al}{\alpha}
\newcommand{\be}{\beta}

\newcommand{\si}{\sigma}
\newcommand{\om}{\omega}
\newcommand{\Om}{\Omega}
\newcommand{\de}{\delta}

\newcommand{\la}{\langle}
\newcommand{\ra}{\rangle}
\newcommand{\tH}{\mathcal{\tilde{H}}}

\newcommand{\dg}{\dagger}
\newcommand{\mH}{\mathcal{H}}
\newcommand{\mV}{\mathcal{V}}
\newcommand{\mS}{\mathcal{S}}

\newcommand{\mG}{\mathcal{G}}
\newcommand{\mZ}{\mathcal{Z}}
\newcommand{\mI}{I}

\def\nn{\nonumber}

\begin{document}
\title{Light propagation through one-dimensional interacting open quantum systems}
\author{Pooja Manasi and Dibyendu Roy}
\affiliation{Raman Research Institute, Bangalore 560080, India}
\begin{abstract}
We apply the quantum Langevin equations approach to study nonlinear light propagation through one-dimensional interacting open quantum lattice models. We write a large set of quantum Langevin equations of lattice operators obtained after integrating out the light fields and use them to derive nonequilibrium features of the lattice models. We first consider a Heisenberg like interacting spin-1/2 chain with nearest-neighbor coupling. The transient and steady-state transport properties of an incoming monochromatic laser light are calculated for this model. We find how the local features of the spin chain and the chain length dependence of light transport coefficient evolve with an increasing power of the incident light. The steady-state light transmission coefficient at a higher power depends non-monotonically on the interaction in a finite chain. While the nonlinear light transmission in our studied model seems to be ballistic in the absence of interaction and for a high interaction, it shows an apparent system-size dependence at intermediate interactions. Later, we extend this method to the long-range interaction between spins of the driven-dissipative lattice model and to incorporate various losses typical in many atomic and solid-state systems.
\end{abstract}

\maketitle

\section{Introduction}

A quantum system can be driven out-of-equilibrium by connecting its two boundaries to different baths which are kept at different chemical potentials, temperatures, densities, pressures, etc. This open quantum system set-up is very popular in the study of nonequilibrium quantum transport \cite{datta_1995, Kohler2005, Nazarov2009}. Connecting a quantum system to baths introduces dissipation and decoherence which play a significant role in thermalization, nonequilibrium phase transition and crossover from quantum to classical transport. Nonequilibrium dynamics in such driven-dissipative systems can be investigated using various theoretical methods including the Keldysh or nonequilibrium Green's function formalism \cite{Kadanoff1962, Keldysh1964, Meir1992, Jauho94, Kamenev2011, Sieberer2016}, the quantum Langevin equations (QLE) approach \cite{Zurcher1990,Segal2003,Kohler2005, DharSen2006,DharRoy2006}, the Lippmann-Schwinger scattering theory \cite{Lippmann1950, Hershfield1993, ShenFanPRL2007} and the Lindblad master equation formalism for the reduced density matrix \cite{Lindblad1976,Gorini1976,Prosen2008,ProsenPRL2011a,ProsenPRL2011b}. Nevertheless, exact results for transient as well as steady-state nonequilibrium dynamics applying these methods are mostly limited to either noninteracting (generally Hamiltonian of the quadratic form in the field operators) quantum systems \cite{Jauho94, Kohler2005, DharSen2006, DharRoy2006} or relatively simple interacting quantum impurity models \cite{Schiller95, Schiro2009, Zheng2010, RoyPRA2014}. Numerical techniques such as real-time Monte Carlo \cite{Muhlbacher2008, Schiro2009, Warner2009}, iterative path-integral \cite{Weiss2008, Segal2010}, time-dependent numerical renormalization group \cite{Anders2008}, time-dependent density-matrix renormalization group (DMRG) \cite{Kirino08, Boulat08, ProzenJSM2009, Longo10, Znidaric2011, Reisons2017, Mascarenhas2017} and time-evolving block decimation (TEBD) \cite{Orus2008} are often used in the study of many-body open quantum systems. However, the constraint on available size in numerics can severely limit the system size and the role of baths in such numerical studies.

 In this paper, we generalize the QLE approach to calculate nonlinear light propagation through one-dimensional (1D) interacting quantum lattice models connected to photon baths at the boundaries. The QLE approach has been previously applied to study nonequilibrium electrical \cite{DharSen2006} and thermal \cite{DharRoy2006} transport in noninteracting and mean-field interacting models \cite{RoyPRB2012}. It can be used both for time-independent and time-dependent transport \cite{Kohler2005}. The QLE approach for nonequilibrium transport is an extension of the Heisenberg-Langevin equation approach \cite{Scully1997, Gardiner2004} to nonequilibrium open quantum systems. Recently, \textcite{RoyPRA2017} applied the QLE approach to study light propagation through a two-level atom (2LA) and a three-level atom embedded in a 1D continuum of photon modes. We here show that the QLE approach can be further extended to efficiently study dynamics of 1D interacting quantum lattice models driven by laser lights. For this, we develop a matrix product operator (MPO) description of writing and solving a large set of quantum Langevin equations of lattice operators which are obtained after integrating out the light fields. We can determine both transient and long-time steady-state properties of the driven-dissipative many-body system precisely. Here, in contrast to many TEBD and time-dependent DMRG studies \cite{Kirino08, Boulat08, Longo10}, we include an infinite number of photon modes in the noninteracting photon baths \footnote{By using normal modes of photons with an infinite bandwidth for the baths and a constant coupling of these modes with the atom, we have missed any non-Markovian effect which can arise in Refs.~\cite{Kirino08, Boulat08, Longo10} due to a finite bandwidth and a real-space coupling.}.  Nevertheless, our current numerical results are restricted to a 1D lattice model of eight sites due to the computation limitation to solving a large set of coupled linear differential equations. 

 We first apply the generalized QLE approach to an atomic medium modeled as a Heisenberg like interacting spin-1/2 chain \cite{Heisenberg1928} with nearest-neighbor coupling. The exact eigenvalues and eigenvectors of the Heisenberg spin-1/2 chain can be derived using the Bethe ansatz \cite{Bethe1931}, and these solutions are crucial in our understanding of equilibrium many-body dynamics in 1D quantum systems \cite{Takahashi1999}. The nonequilibrium dynamics, especially transient dynamics, in an open Heisenberg spin-1/2 chain is complicated to calculate efficiently due to the fast growth of bipartite entanglement and resulting exponential growth of the dimension of the necessary many-body Hilbert space. Nevertheless, some interesting hypotheses have been proposed in the recent years to accurately approximate the nonequilibrium steady-state dynamics of the Heisenberg spin-1/2 chain which is driven out of  equilibrium by Lindblad baths connected at the boundaries of the chain \cite{ProzenJSM2009, ProsenPRL2011a, ProsenPRL2011b, Znidaric2011}. However, the validity of the quantum master equation of Lindblad form in describing nonequilibrium transport in open quantum system of multiple sites is highly contested in the recent years, and these studies seem to suggest the applicability of Lindblad equations can be very restrictive for general baths \cite{Wichterich2007,Levy2014,Purkayastha2016}. Therefore, we here use the QLE approach which has been shown to be exact in the study of nonequilibrium particle and energy transport through a noninteracting open quantum system of multiple sites \cite{DharSen2006, DharRoy2006}. In the following, we also discuss some comparison between the QLE approach and the quantum master equation of Lindblad form in the context of our study. 

 We here find both transient and steady-state transport properties of a monochromatic laser light passing through an atomic medium modeled as a Heisenberg like interacting spin-1/2 chain. The atomic medium is coupled to photon baths at the boundaries. The steady-state laser transmission coefficient in a finite chain depends non-monotonically on the strength of interaction between neighboring atom-photon excitations. We calculate a scaling of the steady-state light transmission with the chain length for the various power of the incident resonant laser beam. While the transmission coefficient seems to show ballistic transport (transmission coefficient is independent of chain length) in the absence of the interaction (equivalent to the XX model) and for a high interaction, it falls with increasing chain length for intermediate interaction strengths at high incident power. We also investigate local properties such as a profile of atomic excitations of the driven-dissipative medium for different laser power. The interacting 1D open quantum media of light have been implemented in many recent experiments with ultra-cold atoms \cite{Peyronel12, Firstenberg13}, ions \cite{Barreiro2011, Schindler2013}, semiconductor qubits \cite{Viennot2014, Viennot2015} and superconductor qubits \cite{vanLoo2013, Fitzpatrick2017}. Motivated by these experimental realizations, we extend the QLE approach to the long-range interaction between atoms of the driven-dissipative lattice model and to incorporate various losses typical in many atomic and solid-state systems \cite{RoyRMP2017}.

The manuscript is organized as follows. In Sec.~\ref{model} we introduce our 1D open quantum optical medium modeled as a nearest-neighbor Heisenberg like interacting spin-1/2 chain connected to photon baths at the boundaries. We describe the generalized QLE approach and related numerical results for a two-site spin chain in Sec.~\ref{2atoms}. In Sec.~\ref{Natoms} we develop the generalized QLE approach using an MPO description for the interacting spin-1/2 chain of many sites. We discuss current challenges and possible future directions for applications of our method in Sec.~\ref{conl}. We also include three appendices: Appendix~\ref{App1} to present single-photon transport in the nearest-neighbor interacting spin-1/2 chain, Appendix~\ref{App2} to incorporate various losses in the spin-1/2 chain, and Appendix~\ref{App3} to extend the current approach to the long-range interaction between spins.

\section{Model}
\label{model}
\begin{figure}
\includegraphics[width=0.99\linewidth]{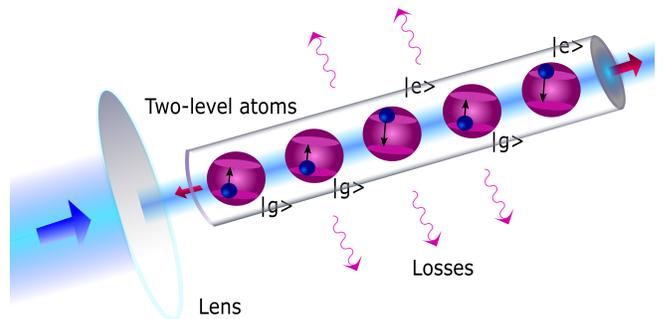}
\caption{A cartoon of a one-dimensional nonlinear quantum optical medium consisting of two-level atoms with levels $|g\ra$ and $|e\ra$. The atomic medium is driven out of equilibrium by a tightly focused, coherent light (blue arrow) shined from the left of the medium. The red arrows show transmitted and reflected lights.}
\label{Cartoon}
\end{figure}

Inspired by recent experimental studies \cite{RoyRMP2017, Peyronel12, Firstenberg13, vanLoo2013, Fitzpatrick2017}, we investigate coherent light propagation through a 1D nonlinear quantum optical medium consisting of real or artificial atoms. The medium is direct-coupled to light fields where the incident light excites (drives) the boundary atoms as in Ref.~\cite{Fitzpatrick2017}, and the atom-photon excitations propagate through the medium due to electromagnetic interactions between the atoms inside the medium (see Fig.~\ref{Cartoon}).  For simplicity, we consider 2LAs with a transition frequency $\om_i$ between ground and excited levels $|g\ra_i$ and $|e\ra_i$ at site $i$. 
We here assume a small separation between the atoms compared to the resonant wavelength of the atoms. This assumption allows us to neglect any non-Markovian features due to retardation of photons between the atoms. First, we develop our theory for the nearest-neighbor interaction between the atoms, and later we extend the theory in Appendix~\ref{App3} to the long-range interaction. The sources of these instantaneous interactions can be various, e.g., mediated by virtual or real photons between atoms in the cavity and circuit quantum electrodynamics (QED) set-ups or dipole-dipole interactions between neutral atoms or Coulomb interactions between ions.

We connect the atomic medium to photonic baths of coherent light at the two boundaries of it to study light propagation through the medium. We consider a linear energy-momentum dispersion $(\omega_k=v_g k)$ for different photon modes in the baths with a group velocity $v_g$ and write atom-photon interactions in linear form within the rotating-wave approximation. The atom-photon coupling strength at the left and right side of the medium are $g_{L}$ and $g_{R}$ respectively. All the couplings are taken in this paper to be constant over photon frequency near the mean $\om_i$; this is known as the Markov approximation causing the photon fields to behave as memoryless baths. 
The total Hamiltonian consisting of the atomic medium of $N$ atoms, the photon baths, and the atom-photon coupling terms is
\bea
\mH_T&=&\mH_M+\mH_{LB}+\mH_{RB}+\mV_{LM}+\mV_{RM}~~{\rm with} \label{Ham}\\
\f{\mH_M}{\hbar}&=&\sum_{i=1}^N \om_i\si_i^\dg\si_i + \sum_{i=1}^{N-1}\big(2J_x(\si_i^\dg\si_{i+1}+\si_{i+1}^\dg\si_i)\nn\\
&&+4J_z\si_i^\dg\si_i\si_{i+1}^\dg\si_{i+1}\big), \label{Hams}\\
\f{\mH_{LB}}{\hbar}&=&\int_{-\infty}^{\infty}dk\:\omega_k a_k^\dg a_k,\f{\mH_{RB}}{\hbar}=\int_{-\infty}^{\infty}dk\:\omega_k b_k^\dg b_k,\\
\f{\mV_{LM}}{\hbar}&=&\int_{-\infty}^{\infty}dk\:g_{L}(a_k^\dg\si_1+\si_1^\dg a_k),\\
\f{\mV_{RM}}{\hbar}&=&\int_{-\infty}^{\infty}dk\:g_{R}(b_k^\dg \si_N+\si_N^\dg b_k).
\eea
Here, $\sigma_i^{\dg}~(\equiv |e\ra_{ii}\la g|)$ and $\sigma_i~(\equiv |g\ra_{ii}\la e|)$ are the raising and lowering operator of the $i$th 2LA, and $a^{\dg}_k, b^{\dg}_k$ create a photon with wave number $k$ respectively at the left and right side photon bath. We set here energy of ground levels $|g\ra_i$ to zero. We have $[a_k,a^{\dg}_{k'}]=[b_k,b^{\dg}_{k'}]=\delta(k-k')$ and $\sigma_j^{\dg}=(\sigma_j^{x}+i\sigma_j^{y})/2,~\sigma_j^{z}=2\sigma_j^{\dg}\sigma_j-\mI_j$ where $\sigma_j^{x}, \sigma_j^{y}$ and $\sigma_j^{z}$ are $2\times2$ Pauli matrices with commutation relation $[\sigma_j^{x}, \sigma_k^{y}]=2i\delta_{jk}\sigma_j^{z}$ and $\mI_j$ is a $2\times2$ unit matrix in local Hilbert space of the $j$th atom.

For the coupling constants $J_{x}=J_{z}$, the Hamiltonian in Eq.~\ref{Hams} of the atomic medium is a bit similar to the isotropic Heisenberg model of spin-1/2 in external magnetic fields where $\om_i$'s simulate external magnetic fields \footnote{When we transform the Pauli matrices by the ladder operators in the spin-1/2 Heisenberg model, the local fields in transformed Hamiltonian are $\om_{sj}=\om_j-2J_z$ for $j=1,N$ and $\om_{sj}=\om_j-4J_z$ for $j=2,3,\dots N-1$. This rescaling of local fields or transition frequencies creates inhomogeneties at the boundaries which have significant consequences for shorter chain lengths. Here we ignore such rescaling of the  transition frequencies for simplicity.}. When $J_{x} \ne J_{z}$, this is like the Heisenberg XXZ model in external fields. The Heisenberg spin-1/2 chain can be mapped to a 1D Hubbard model of interacting spinless fermions via the Jordon-Wigner transformation. For $J_{x} \ne 0$ and $J_{z}=0$, the medium Hamiltonian is the XX spin chain which transforms into a tight-binding chain of free spinless fermions.

It has been recently argued from perturbative calculation and simulation that the linear-response spin transport is ballistic and superdiffusive anomalous, respectively, for $J_z<J_x$ and $J_z=J_x$ (isotropic) in the Heisenberg spin chain in the absence of external magnetic field at any temperature \cite{ProzenJSM2009, Znidaric2011}. According to these studies, the linear response spin current is expected to be diffusive at infinite temperature for $J_z>J_x$. On the other hand, another study \cite{ProsenPRL2011b} based on the Lindblad equation claims that the nonequilibrium spin current is independent of chain length $N$ (ballistic) for $J_z<J_x$, the spin current scales as $1/N^2$ for $J_z=J_x$, and it decays exponentially with $N$ (insulator) for $J_z>J_x$. We find here that the nonequilibrium (nonlinear) transmission of resonant laser light is ballistic for $J_z\ll J_x$ and $J_z\gg J_x$, and it shows a clear $N$-dependence when $J_z=J_x$. The above results show that the nonequilibrium spin and photon transport can be very different in the Heisenberg spin chain when $J_z>J_x$.


A single photon saturates a single 2LA, and the 2LA acts as a nonlinear medium for two or more photons. Apart from this optical nonlinearity at individual atoms, the $J_z$ coupling can also induce optical nonlinearity via interaction between atom-photon excitations. Therefore, two different mechanisms are responsible for optical nonlinearity in our model. Our photon baths here are kept at zero temperature. These baths act as a source of coherent lights which can propagate through the atomic medium by creating atom-photon excitations. In this paper, we study the propagation of these lights and the consequent radiative energy transfer by them through the medium.    

We assume here that the atom-photon couplings are turned on at $t=t_0$ when a light beam is shined on the atomic medium. In the following, we first formulate our method to study light propagation in a small atomic medium of two atoms. It can mostly be carried out analytically. Later, we extend the technique to a more extended medium of $N$ atoms using MPOs.  

\section{Nonlinear medium of two atoms}
\label{2atoms}
The single-photon transport properties in an atomic chain of $N$ atoms can be calculated following \textcite{RoyNat2013}. The transmission and reflection coefficients are extracted from the single-photon scattering eigenstate on real-space which we derive in Appendix \ref{App1}. For example, the single-photon transmission amplitude $t_2(\om_p)$ of an incident photon at frequency $\om_p$ on a medium of two atoms is
\bea
t_{2}(\om_p)=\f{-4iJ_x \sqrt{\Gamma_L\Gamma_R}}{(\om_p-\om_1+i\Gamma_L)(\om_p-\om_2+i\Gamma_R)-4J_x^2},\label{1trans}
\eea
where the atom-photon coupling rates, $\Gamma_L=\pi g_L^2/v_g$ and $\Gamma_R=\pi g_R^2/v_g$. The transmission coefficient $|t_2(\om_p)|^2$ shows two resonant peaks around photon frequency $\om_{p\pm}=(\om_1+\om_2)/2 \pm \sqrt{(\om_1-\om_2)^2+(4J_x^2+\Gamma_L\Gamma_R)}$. These peaks are due to the resonant exchange of photon between two atoms. The above expression for $t_2(\om_p)$ is independent of $J_z$ which causes an effective interaction between two atom-photon excitations and is absence for single atom-photon excitation. We now examine how the transmission lineshape of a laser beam changes as we increase the power of the incident laser beam from a very low (single-photon limit) to a high value.  

We calculate nonlinear light propagation through the medium using the QLE approach. We start the calculation by writing the Heisenberg equations  for photonic operators of the baths and atomic operators of the medium. One important feature of the interaction between atoms is that the  Heisenberg equations for individual (local) atomic operators ($\sigma_i,\sigma^{\dg}_i,\sigma^{\dg}_i\sigma_i$) do not form a close set as these equations have terms with higher-order (non-local) operators involving different atoms. Therefore, we also need to write Heisenberg equations for these higher-order atomic operators involving different atoms. For example, we need to consider all the atomic operators in $\{\mI_1,\si_1^\dg,\si_1,\si_1^\dg\si_1\}\otimes \{\mI_2,\si_2^\dg,\si_2,\si_2^\dg\si_2\}$ for two atoms. For an atomic chain of $N$ atoms, the complete set of atomic operators to form a close set of Heisenberg equations is $\{\mI,\si^\dg,\si,\si^\dg\si\}^{\otimes N}$. Therefore, we have $4^N-1$ nontrivial Heisenberg equations for $N$ atoms as $\mI^{\otimes N}$ has no dynamics.

The Heisenberg equations for the photon operators $a_k,b_k$ are first-order linear inhomogeneous differential equations which we solve formally for some initial condition at $t_0$. The initial condition of photon operators provides a direction of incoming light. For the atomic medium of two atoms, we get time-evolution of the photon operators, $a_k(t)$ and $b_k(t)$ with initial conditions $a_k(t_0)$ and $b_k(t_0)$ as 
\bea
a_k(t)&=&\mG_k(t-t_0)a_k(t_0)-ig_L \int_{t_0}^{t} dt'\mG_k(t-t') \sigma_1(t'),\label{HEsol1} \\
b_k(t)&=&\mG_k(t-t_0)b_k(t_0)-ig_R \int_{t_0}^{t} dt'\mG_k(t-t') \sigma_2(t'),\label{HEsol2}
\eea
with $\mG_k(\tau)=e^{-iv_gk\tau}$. Plugging these solutions of the photon operators in the Heisenberg equations of the  atomic operators $\si_1, \si_1^\dg\si_1, \si_2, \si_2^\dg\si_2, \si_1\si_2, \si_1^{\dg}\si_2, \si_1 \si_2^\dg\si_2, \si_1^\dg\si_1\si_2, \si_1^\dg\si_1\si_2^\dg\si_2$ and their nontrivial hermitian conjugates, 
we derive a set of (fifteen) nonlinear QLE of the atomic operators \cite{RoyPRA2017}. The noises in these Langevin equations are coming from the left and right photon baths and we denote them here by $\eta_L(t)=\int_{-\infty}^{\infty}dk\:\mG_k(t-t_0)g_La_k(t_0)$ and $\eta_R(t)=\int_{-\infty}^{\infty}dk\:\mG_k(t-t_0)g_Rb_k(t_0)$. The properties of these noises are determined by the initial condition of the photon fields at $t=t_0$. For example, for an incident laser light in the coherent state $|E_p,\om_p\ra$ with a frequency $\om_p$ and an amplitude $E_p$ from the left of the atomic medium, we have
\bea
a_k(t_0)|E_p,\om_p\ra &=& E_p\:\delta(v_gk-\om_p)|E_p,\om_p\ra, \label{ini1}\\
b_k(t_0)|E_p,\om_p\ra &=& 0. \label{ini2}
\eea
We can also have incoming light from the both sides of the atomic medium. 

To solve the nonlinear QLE of atomic operators, we apply the above properties of incident light in the coherent state. Thus, we transform the nonlinear operator equations into a set of linear coupled differential equations of non-operator variables by performing expectation of these operator-equations in the initial state $|E_p,\om_p\ra$. Taking expectation also converts the noise operators in these QLE to c-numbers. We remove any explicit time-dependence in the coefficients of the linear coupled differential equations of the non-operator variables by introducing the following variables with $j=1,2$: \cite{RoyPRA2017, Koshino12}
\bea
\mathcal{S}_j(t)&=&\la \sigma_j(t) \ra e^{i\om_p(t-t_0)},\label{s1}\\
\mathcal{S}_{jj}(t)&=& \la \sigma_{j}^{\dg}(t)\sigma_{j}(t)\ra,\label{s2} \\
\mathcal{S}_{3}(t)&=& \la \sigma_{1}(t)\sigma_{2}(t)\ra e^{2i\om_p(t-t_0)},\label{s3} \\
\mathcal{S}_{12}(t)&=& \la \sigma_{1}^{\dg}(t)\sigma_{2}(t)\ra,\label{s4} \\
\mathcal{S}_{122}(t)&=& \la \sigma_{1}(t)\sigma_{2}^{\dg}(t)\sigma_{2}(t)\ra e^{i\om_p(t-t_0)},\label{s5} \\
\mathcal{S}_{112}(t)&=& \la \sigma_{1}^{\dg}(t)\sigma_{1}(t)\sigma_{2}(t)\ra e^{i\om_p(t-t_0)},\label{s6}\\
\mathcal{S}_{1122}(t)&=& \la \sigma_{1}^{\dg}(t)\sigma_{1}(t) \sigma_{2}^{\dg}(t)\sigma_{2}(t)\ra \label{s7},
\eea
where we multiply a factor $e^{i\om_p(t-t_0)}$ with each $\si_j(t)$ and $e^{-i\om_p(t-t_0)}$ with each $\si_j^\dg(t)$ of the atomic operators in $\{\mI_1,\si_1^\dg,\si_1,\si_1^\dg\si_1\}\otimes \{\mI_2,\si_2^\dg,\si_2,\si_2^\dg\si_2\}$. Here we use the following convention for writing the expectation of an operator $\hat{O}$ in $|E_p,\om_p\ra$, $\la E_p,\om_p|\hat{O}|E_p,\om_p\ra \equiv \la \hat{O}\ra$. Thus, we obtain the following differential equations for the above variables in Eqs.~\ref{s1}-\ref{s7}:
\bea
\f{d\mathcal{S}_{1}}{dt}&=&-\big(i \delta \om_1+\Gamma_L\big)\mathcal{S}_{1}+2iJ_x(2\mathcal{S}_{112}-\mathcal{S}_{2})-4iJ_z\mathcal{S}_{122}\nn\\&&+2i\Omega_L\mathcal{S}_{11}-i\Omega_L,\label{s8} \\
\f{d\mathcal{S}_{2}}{dt}&=&-\big(i \delta \om_2+\Gamma_R\big)\mathcal{S}_{2}+2iJ_x(2\mathcal{S}_{122}-\mathcal{S}_{1})-4iJ_z\mathcal{S}_{112}, \nn\\ \label{s9} \\
\f{d\mathcal{S}_{11}}{dt}&=&-2\Gamma_L\mathcal{S}_{11}-2iJ_x(\mathcal{S}_{12}-\mathcal{S}^*_{12})+i\Omega_L(\mathcal{S}_{1}-\mathcal{S}^*_{1}),\nn\\ \label{s10} \\
\f{d\mathcal{S}_{22}}{dt}&=&-2\Gamma_R\mathcal{S}_{22}+2iJ_x(\mathcal{S}_{12}-\mathcal{S}^*_{12}),\label{s11}\\
\f{d\mathcal{S}_{3}}{dt}&=&-\big(i(\delta \om_1+\delta \om_2+4J_z)+\Gamma_L+\Gamma_R\big)\mathcal{S}_{3}\nn\\&&+i\Omega_L(2\mathcal{S}_{112}-\mathcal{S}_{2}), \label{s12}
\eea
\bea
\f{d\mathcal{S}_{12}}{dt}&=&\big(i(\om_1-\om_2)-\Gamma_L-\Gamma_R\big)\mathcal{S}_{12}+2iJ_x(\mathcal{S}_{22}-\mathcal{S}_{11})\nn\\&&+i\Omega_L(\mathcal{S}_{2}-2\mathcal{S}_{112}), \label{s13} \\
\f{d\mathcal{S}_{122}}{dt}&=&-\big(i(\delta \om_1+4J_z)+\Gamma_L+2\Gamma_R\big)\mathcal{S}_{122}+2iJ_x\mathcal{S}_{112}\nn\\&&+i\Omega_L(2\mathcal{S}_{1122}-\mathcal{S}_{22}),\label{s14}\\
\f{d\mathcal{S}_{112}}{dt}&=&-\big(i(\delta \om_2+4J_z)+2\Gamma_L+\Gamma_R\big)\mathcal{S}_{112}+2iJ_x\mathcal{S}_{122}\nn\\&&+i\Omega_L(\mathcal{S}_{3}-\mathcal{S}_{12}), \label{s15}\\
\f{d\mathcal{S}_{1122}}{dt}&=&-2(\Gamma_L+\Gamma_R)\mathcal{S}_{1122}+i\Omega_L(\mathcal{S}_{122}-\mathcal{S}^*_{122}), \label{s16}
\eea
where the detuning $\delta \om_j=\om_j-\om_p$ with $j=1,2$ and the Rabi frequency of the incident laser beam, $\Omega_L=g_LE_p/v_g$. The rates $\Gamma_L,\Gamma_R$ denote dissipation in the atomic medium due to its coupling to the photon baths at the left and right boundaries. The rest of the nontrivial equations in the complete set of fifteen coupled differential equations are complex conjugates of the equations in \ref{s8},\ref{s9},\ref{s12},\ref{s13},\ref{s14},\ref{s15}. The Eqs.~\ref{s8}-\ref{s16} obtained after averaging over the light fields in the coherent state are the optical Bloch equations or the master equations of Lindblad form written in a specific basis \cite{Cohen1992atom,Daley2014}.

The equations \ref{s8}-\ref{s16} can be expressed in a compact manner by introducing the vectors $\boldsymbol{\mathcal{S}}_2=(\mathcal{S}_1^*,\mathcal{S}_1,\mathcal{S}_{11}$, $\mathcal{S}_2^*,\mathcal{S}_3^*,\mathcal{S}_{12}^*,\mathcal{S}^*_{112},\mathcal{S}_2,\mathcal{S}_{12},\mathcal{S}_3,\mathcal{S}_{112},\mathcal{S}_{22},\mathcal{S}_{122}^*,\mathcal{S}_{122},\mathcal{S}_{1122})^{T}$ and $\boldsymbol{\Omega}_2=$ $(i\Omega_L,-i\Omega_L, 0,0,0,0,0,0,0,0,0,0,0,0,0)^{T}$: 
\bea
\f{d\boldsymbol{\mathcal{S}}_2}{dt}=\boldsymbol{\mathcal{Z}}_2\boldsymbol{\mathcal{S}}_2+\boldsymbol{\Omega}_2,\label{sigma5}
\eea
where $\boldsymbol{\mathcal{Z}}_2$ includes the matrix elements of the Hamiltonian $\mH_M$ plus the driving terms due to the laser field and the dissipation due to coupling to the baths.
 The Eq.~\ref{sigma5} for such non-operator variables can be solved with an initial condition, e.g., $\boldsymbol{\mathcal{S}}_2(t=t_0)=0$ which indicates the atoms in the ground state before shining a light on the medium. 
The long-time steady-state behavior of the medium is independent of the initial condition of the atomic chain. The  steady-state solutions are obtained by setting $\f{d\boldsymbol{\mathcal{S}}_2}{dt}=0$. Therefore, we have at the steady-state, $\boldsymbol{\mathcal{S}}_2(t=\infty)=-\boldsymbol{\mathcal{Z}}_2^{-1}\boldsymbol{\Omega}_2$ which involves the inverse of a fifteen dimensional square matrix. We invert this matrix numerically.

The transmission and reflection coefficients of the laser beam can be calculated from the scattered power. For this, we introduce a real-space description of the photon operators at position $x \in [-\infty,\infty]$ of both sides of the atomic medium \cite{RoyPRA2017}. For a linear energy-momentum dispersion of photons in the baths, we define photon operator as $a_x(t)=\int_{-\infty}^{\infty} dk\: e^{ikx}a_k(t)/\sqrt{2\pi}$ and $b_x(t)=\int_{-\infty}^{\infty} dk\: e^{ikx}b_k(t)/\sqrt{2\pi}$ where the operators at $x<0$ and $x>0$ represent respectively incident and scattered photons on each side of the atomic medium and the photons at $x=0$ are coupled to the atom at the boundaries of the medium.

For an incident light from left of the medium, the power spectrum of the incident light is defined as
\bea
P_{\rm in}(\om)={\rm Re}\int_0^{\infty}\f{d\tau}{\pi}e^{i\om \tau}\la a_{x<0}^{\dg}(t)a_{x<0}(t+\tau)\ra,\label{powerIn} 
\eea
where we take $t<t_0$ and the expectation $\la..\ra$ is again performed in the initial state $|E_p,\om_p\ra$. We find $P_{\rm in}(\om)=E_p^2\:\delta(\om-\om_p)/(2\pi v_g^2)$. The intensity (total number of photons) of the incident light per unit quantization length, $\int d\om P_{\rm in}(\om)=E_p^2/(2\pi v_g^2)=I_{\rm in}$. 
 The power spectrum of the transmitted and the reflected light are defined, respectively, as
\bea
P_{\rm tr}(t,\om)&=&{\rm Re}\int_0^{\infty}\f{d\tau}{\pi}e^{i\om \tau}\la b_{x>0}^{\dg}(t)b_{x>0}(t+\tau)\ra,\label{powerTr} \\
P_{\rm rf}(t,\om)&=&{\rm Re}\int_0^{\infty}\f{d\tau}{\pi}e^{i\om \tau}\la a_{x>0}^{\dg}(t)a_{x>0}(t+\tau)\ra,\label{powerTr}
\eea
where $t>t_0$. $P_{\rm tr}(t,\om)$ and $P_{\rm rf}(t,\om)$ become independent of time $t$ at a long-time steady-state. To calculate $P_{\rm tr}(t,\om)$ and $P_{\rm rf}(t,\om)$, we first apply the formal solution of the Heisenberg equation for $b_k(t)$ and $a_k(t)$ from Eqs.~\ref{HEsol1},\ref{HEsol2}, and rewrite  $P_{\rm tr}(t,\om)$ and $P_{\rm rf}(t,\om)$ using input fields and atomic operators. Applying Eqs.~\ref{ini1},\ref{ini2}, we find
\bea
P_{\rm tr}(t,\om)&=&\f{2\Gamma_R}{\pi v_g}{\rm Re}\int_0^{\infty}d\tau e^{i\om \tau} \la \sigma_2^{\dg}(t)\sigma_2(t+\tau)\ra,\\
P_{\rm rf}(t,\om)&=&P_{\rm in}(\om)-\f{\Omega_L}{\pi v_g}{\rm Im}\int_0^{\infty}d\tau e^{i(\om-\om_p)\tau} \nn\\&&\times \big( \mathcal{S}_1^*(t)-\mathcal{S}_1(t+\tau)\big)\nn\\&&+\f{2\Gamma_L}{\pi v_g}{\rm Re}\int_0^{\infty}d\tau e^{i\om \tau} \la \sigma_1^{\dg}(t)\sigma_1(t+\tau)\ra,
\eea
where we take $x \to 0+$. Thus, one needs to calculate a two-time correlation of atomic operators $\la \sigma_1^{\dg}(t)\sigma_1(t+\tau)\ra$ and $\la \sigma_2^{\dg}(t)\sigma_2(t+\tau)\ra$ to find the transmitted and reflected power spectra which are useful to derive transmission and reflection of radiation energy. The two-time correlations can be evaluated at steady-state by deriving a set of coupled differential equations for the two-time correlation of atomic operators which are similar in the form of Eqs.~\ref{s8}-\ref{s16}. It has been carried out in Ref.~\cite{RoyPRA2017} for a single atom. However, we are here only interested in getting total transmitted and reflected power which in turn would give transmission and reflection coefficients of light. The time-dependent transmission and reflection coefficient for an incoming light from the left of the medium are respectively,
\bea
\mathcal{T}_2(t)&=&\int d\omega P_{\rm tr}(t,\omega)/I_{\rm in}\nn\\&=&\f{2\Gamma_R}{v_gI_{\rm in}} \langle \sigma_2^{\dg}(t)\sigma_2(t) \rangle=\f{2\Gamma_R}{v_gI_{\rm in}}\mathcal{S}_{22}(t),\label{trans} \\
\mathcal{R}_2(t)&=&\int d\omega P_{\rm rf}(t,\omega)/I_{\rm in}\nn\\&=&1+\f{2\Omega_L}{v_gI_{\rm in}}{\rm Im}[\mathcal{S}_1(t)]+\f{2\Gamma_L}{v_gI_{\rm in}} \mathcal{S}_{11}(t).\label{refl}
\eea

\begin{figure}
\includegraphics[width=0.99\linewidth]{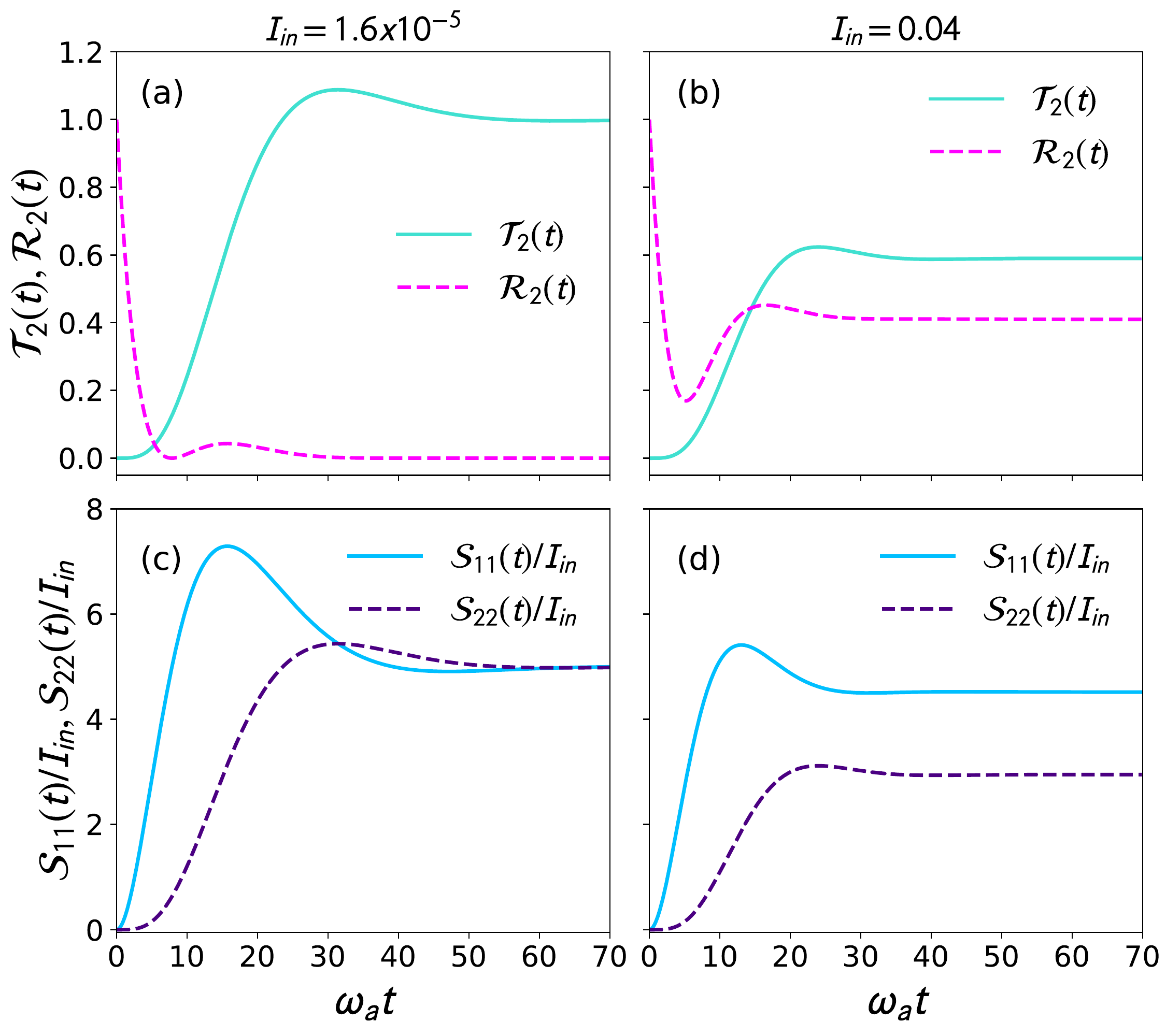}
\caption{Transient properties of the scattered light and the atoms in a medium of two atoms driven by a laser light. The upper row shows time-evolution of transmission and reflection coefficients $\mathcal{T}_2(t), \mathcal{R}_2(t)$ of a laser light from the left of the medium modeled as an interacting spin chain. The lower row depicts time-evolution of excited atoms scaled by power of the incident laser. The incident laser power in the first and second column, respectively, is $I_{\rm in}=1.6 \times 10^{-5}$ and $0.04$ (in units of $\om_a/v_g$). The other parameters are $\om_1=\om_2=\om_p=\om_a, J_x=J_z=0.05\om_a, \Gamma_L=\Gamma_R=0.1\om_a$.}
\label{transient}
\end{figure}

\begin{figure}
\includegraphics[width=0.99\linewidth]{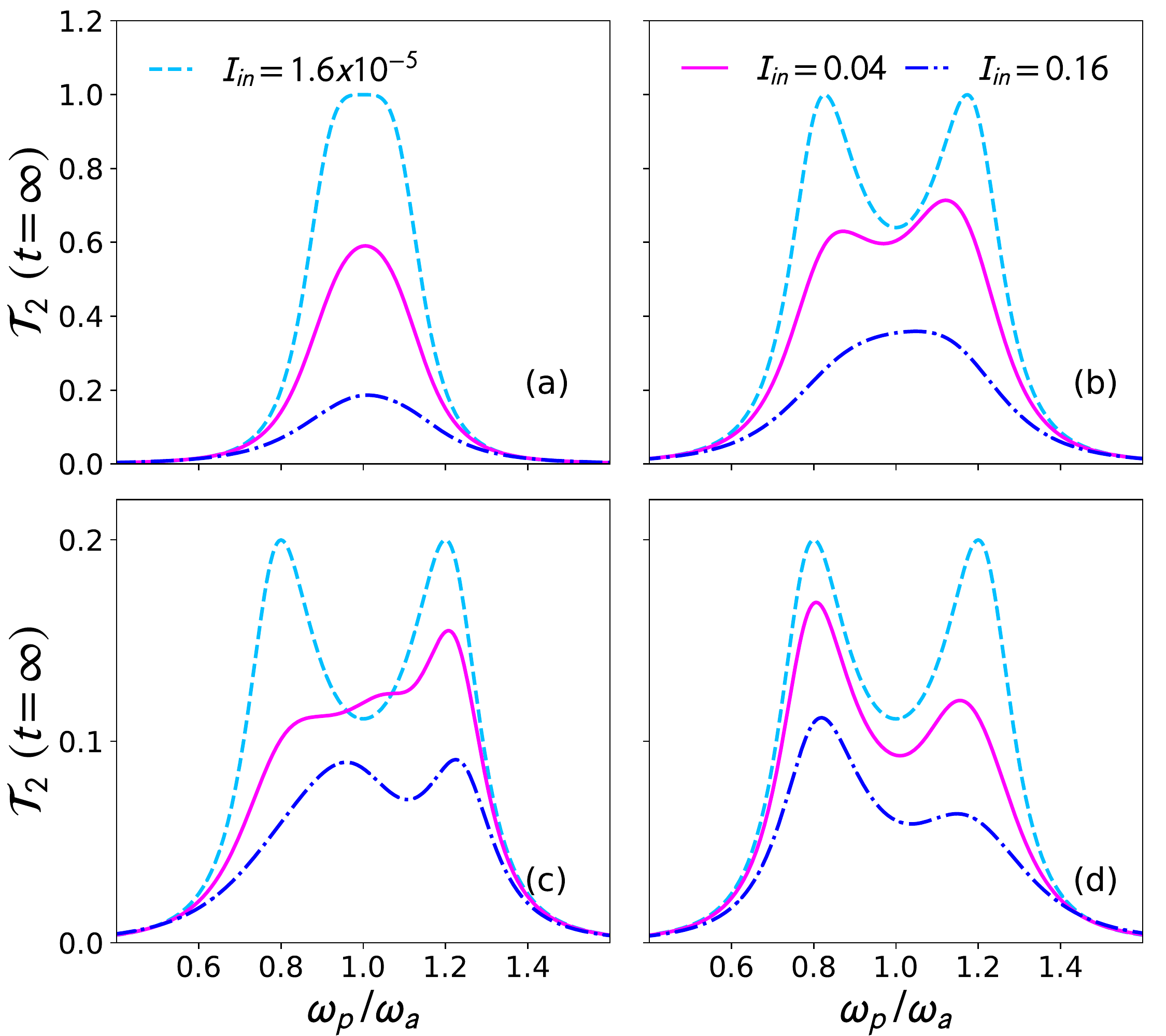}
\caption{Linear and nonlinear laser transmission through an atomic medium of two atoms modeled as an interacting spin chain. The steady-state transmission coefficient $\mathcal{T}_2(t=\infty)$ vs. scaled frequency $\omega_p/\omega_a$ of the laser for various intensities ($I_{\rm in} \propto E_p^2$) of the incident laser. The single-photon (linear) transmission $|t_2(\omega_p)|^2$ in Eq.~\ref{1trans} is identical to $\mathcal{T}_2(t=\infty)$ for $I_{\rm in}=1.6\times 10^{-5}$. The parameters are (a) $\om_1=\om_2=1, J_x=0.05$, (b) $\om_1=\om_2=1,J_x=0.1$, (c) $\om_1=0.8,\om_2=1.2,J_x=0.05$, and (d) $\om_1=1.2,\om_2=0.8,J_x=0.05$. In all the plots $\Gamma_L=\Gamma_R=0.1,J_z=0.05$. The rates $\Gamma_L,\Gamma_R, J_x, J_z$ and $\om_1,\om_2$ are in units of $\om_a$, and $I_{\rm in}$ is in units of $\om_a/v_g$.}
\label{twoatomsa}
\end{figure}

{\it Numerical results}: Next, we present some numerical results for the transient and steady-state properties of the scattered light as well as of the atomic medium driven by a laser light from the left. We set $v_g=1$ in all our numerical results. We solve  Eq.~\ref{sigma5} with the initial condition, $\boldsymbol{\mathcal{S}}_2(t_0=0)=0$ which describes all atoms in the ground state before shining the laser on them. Using the solution of $\boldsymbol{\mathcal{S}}_2(t)$, we evaluate $\mathcal{T}_2(t), \mathcal{R}_2(t)$ of the incident laser light from the left of the medium modeled as an interacting spin chain. We show $\mathcal{T}_2(t), \mathcal{R}_2(t)$ with time in Figs.~\ref{transient}(a) and \ref{transient}(b) for a resonant incident laser ($\om_p=\om_1=\om_2$) at a lower $(I_{\rm in}=1.6 \times 10^{-5},~E_p=0.01)$ and a higher $(I_{\rm in}=0.04,~E_p=0.5)$ power respectively. 

For a very weak incident laser (single-photon limit), we find $\mathcal{T}_2(t)$ grows from zero at the initial time as the laser passes through the atoms. However, $\mathcal{T}_2(t)$ exceeds one for a short time window before saturating to near one as shown in Fig.~\ref{transient}(a) for $I_{\rm in}=1.6 \times 10^{-5}$. According to our definition of the reflection coefficient, it is one when there is no interaction between light and the medium. Therefore, $\mathcal{R}_2(t)$ starts from one at $t=t_0=0$ here, and it decays to zero as the laser passes through the medium. However, $\mathcal{R}_2(t)$ shows a bump at some intermediate time before saturating at zero. The appearance of $\mathcal{T}_2(t)>1$ and a bump in $\mathcal{R}_2(t)$ at some intermediate time can be related to the initial trapping of photons in the medium as atomic excitations. These atomic excitations are stimulated by the incident light to emit extra photons in the forward and backward directions. This stimulated emission generates a temporary amplification of transmission and reflection coefficient. The above phenomenon is also the reason for the humps in the time-evolution of scaled atomic excitation amplitudes $\mathcal{S}_{11}(t)/I_{\rm in}, \mathcal{S}_{22}(t)/I_{\rm in}$ shown in Fig.~\ref{transient}(c). For a weak resonant laser and $2J_x=\Gamma_L=\Gamma_R$, both the atoms are equally excited at long-time steady state. The above features also remain the same for an atomic medium modeled as an XX spin chain $(J_z=0)$ as the medium is mostly linear at such a weak laser power.

For a higher laser power $(I_{\rm in}=0.04)$, many above details of the transport coefficients and the atomic excitations for a lower power are changed due to the optical nonlinearity of the atoms at the high power. While $\mathcal{T}_2(t)$ does not approach one, $\mathcal{R}_2(t)$ never vanishes at a higher laser power (see Fig.~\ref{transient}(b)). We also find in Fig.~\ref{transient}(d) that the steady-state atomic excitation amplitudes are a bit different from Fig.~\ref{transient}(c). We have also checked that though the transient behavior of $\mathcal{T}_2(t), \mathcal{R}_2(t), \mathcal{S}_{11}(t)/I_{\rm in}, \mathcal{S}_{22}(t)/I_{\rm in}$ are very different for different $\boldsymbol{\mathcal{S}}_2(t_0)$, the long-time steady-state properties of the above quantities are independent of $\boldsymbol{\mathcal{S}}_2(t_0)$ as implied by Eq.~\ref{sigma5}. 

The long-time steady-state transmission coefficient $\mathcal{T}_2(t=\infty)$ of an incoming laser light from the left of the atomic  medium is shown Fig.~\ref{twoatomsa}. We plot $\mathcal{T}_2(t=\infty)$ with the frequency $\omega_p$ of the incident light in Fig.~\ref{twoatomsa} for various intensities  $I_{\rm in}~(\propto E_p^2)$ of the laser light and different parameters of the atomic medium. The lineshape of $\mathcal{T}_2(t=\infty)$ matches with the single-photon transmission $|t_2(\omega_p)|^2$ in Eq.~\ref{1trans} when the power of the incident laser is very low, i.e., $\Omega_{L}<<\Gamma_L,\Gamma_R$. The solid blue lines in Fig.~\ref{twoatomsa} for $\Omega_L=0.0018~ (I_{\rm in}=1.6 \times 10^{-5},~E_p=0.01)$ are identical to the single-photon transmission  $|t_2(\omega_p)|^2$ through the atomic medium. As discussed before, there would be two resonant peaks in $|t_2(\omega_p)|^2$ lineshape due to the resonant exchange of photon by the coupling $J_x$ between two atoms. However, for identical atoms when $2J_x \le \Gamma_L,\Gamma_R$, these resonant peaks overlap with each other to become one due to the broadening of the individual peaks by the bath couplings $\Gamma_L,\Gamma_R$. It is the case in Fig.~\ref{twoatomsa}(a). The separation between the resonant peaks in $\mathcal{T}_2(t=\infty)$ at low power ($I_{\rm in}=1.6 \times 10^{-5},~E_p=0.01$) is evident in Fig.~\ref{twoatomsa}(b) for a higher $2J_x>\Gamma_L,\Gamma_R$. It is also clearly visible in Figs.~\ref{twoatomsa}(c,d) even for $2J_x \le \Gamma_L,\Gamma_R$ when the transition frequencies of the atoms are very different. Note that, the maximum value of $\mathcal{T}_2(t=\infty)$ for two different atoms in Figs.~\ref{twoatomsa}(c,d) is much smaller than that in Figs.~\ref{twoatomsa}(a,b) for the identical atoms. This is because light transmission falls rapidly with increasing detuning of light's frequency from atomic transition frequencies.

The magnitude of $\mathcal{T}_2(t=\infty)$ decreases from $|t_2(\omega_p)|^2$ with an increasing power (higher $E_p$) of the incident laser as shown in all the plots of Fig.~\ref{twoatomsa}. The primary reason is the optical nonlinearity of the 2LAs. 
This optical nonlinearity results in photon blockade which is similar to Coulomb blockade in electrical transport. Secondarily, the contribution of instantaneous interaction $J_x$ between the atoms is maximum when one atom is in the excited state while the other is in the ground state. This is the case for a single-photon incoming light. As the photon number or the power of the incident light increases, there is a higher probability for both the atoms to be simultaneously excited. Therefore, the contribution of the $J_x$ interaction in light transmission reduces with increasing intensity of light, and the transmission $\mathcal{T}_2(t=\infty)$ falls. It is also the reason for decreasing separation between the resonant peaks (created by the $J_x$ interaction) which eventually become one around the atoms' transition frequency at a higher intensity as shown in Fig.~\ref{twoatomsa}(b).

Finally, we point out that while the lineshape of $\mathcal{T}_2(t=\infty)$ with $\omega_p$ in Fig.~\ref{twoatomsa} is symmetric across the mean atomic transition frequency $(\om_1+\om_2)/2$ for the single-photon incoming light; it becomes asymmetric for higher incident light power in Figs.~\ref{twoatomsa}(b,c,d). The asymmetry is created by an intensity-dependent nonlinear contribution to the refractive index which is also spatially asymmetric for different transition frequencies $(\om_1\ne \om_2)$ of the 2LAs or/and when $\Gamma_L \ne \Gamma_R$. For two identical atoms and $\Gamma_L=\Gamma_R$, the asymmetry of nonlinear $\mathcal{T}_2(t=\infty)$ around the mean frequency is absent when $J_z=0$. This spatially asymmetric and nonlinear refractive index is responsible for nonreciprocity in light propagation through such spatially asymmetric nonlinear media \cite{RoyNat2013, FratiniPRL2014}. 

\section{Nonlinear medium of many atoms}
\label{Natoms}
We now consider an atomic medium of several atoms. Here we generalize the above QLE approach to study linear and nonlinear light transmission through a long atomic chain. In general, it is challenging to write and solve a large set of $(4^N-1)$ QLE for the atomic medium of $N$ atoms. To do this, we have developed a numerical technique using an MPO description of the Hamiltonian.

The primary step of the numerical method is to efficiently calculate the commutation relation between the atomic operators and the total Hamiltonian for the Heisenberg equations. The atomic operators are written as a tensor product of local operators: $\{\mI_1,\si_1^\dg,\si_1,\si_1^\dg\si_1\}\otimes \{\mI_2,\si_2^\dg,\si_2,\si_2^\dg\si_2\}\otimes \dots \otimes \{\mI_N,\si_N^\dg,\si_N,\si_N^\dg\si_N\} \equiv \boldsymbol{\sigma}_1\otimes \boldsymbol{\sigma}_2 \otimes  \dots \otimes \boldsymbol{\sigma}_N\equiv \boldsymbol{\sigma}^{\otimes N}$. Following the calculation in Sec.~\ref{2atoms} for two atoms, we multiply a factor $e^{i\om_p(t-t_0)}$ with each $\si_j(t)$ and $e^{-i\om_p(t-t_0)}$ with each $\si_j^\dg(t)$ of the atomic operators in $\boldsymbol{\sigma}^{\otimes N}$. After multiplication of these factors, we call the atomic operators as $\tilde{\boldsymbol{\sigma}}^{\otimes N}$. This redefinition of atomic operators as $\tilde{\boldsymbol{\sigma}}^{\otimes N}$ removes any explicit time-dependence in the coefficients of linear coupled differential equations of the non-operator variables $\la \tilde{\boldsymbol{\sigma}}^{\otimes N}\ra$ obtained after taking expectation of the Heisenberg equations of $\tilde{\boldsymbol{\sigma}}^{\otimes N}$ in the initial state $|E_p,\om_p\ra$ as in Sec.~\ref{2atoms}. The Heisenberg equations of $\tilde{\boldsymbol{\sigma}}^{\otimes N}$ are:
\bea
\f{d\tilde{\boldsymbol{\sigma}}^{\otimes N}}{dt}=-\f{i}{\hbar}[\tilde{\boldsymbol{\sigma}}^{\otimes N},\mH_T]+\f{\partial\tilde{\boldsymbol{\sigma}}^{\otimes N}}{\partial t},\label{HEq}
\eea
where the last term is due to explicit time-dependence of $\tilde{\boldsymbol{\sigma}}^{\otimes N}$. We get rid of the term $\partial\tilde{\boldsymbol{\sigma}}^{\otimes N}/(\partial t)$ from the Heisenberg equations of $\tilde{\boldsymbol{\sigma}}^{\otimes N}$ by replacing $\om_j$ in $\mH_M$ by $\delta \om_j=\om_j-\om_p$. After replacing $\om_j$ by $\delta \om_j$ in $\mH_M$ and $\mH_T$, we rename them as $\tilde{\mH}_M$ and $\tilde{\mH}_T$ respectively. Therefore, we can now write Eq.~\ref{HEq} as
\bea
\f{d\tilde{\boldsymbol{\sigma}}^{\otimes N}}{dt}=-\f{i}{\hbar}[\tilde{\boldsymbol{\sigma}}^{\otimes N},\tilde{\mH}_T].\label{HEq1}
\eea

We rewrite the renormalized total Hamiltonian as $\tilde{\mH}_T=\tilde{\mH}_M+\mH_{E}$ where $\mH_{E}$ includes the Hamiltonian of the photon baths and the atom-photon couplings. For nearest-neighbor interaction between atoms, the Hamiltonian  of the medium can be expressed as a sum over nearest-neighbor bond Hamiltonians:
\bea
\tilde{\mH}_M&=&\sum_{i=1}^{N-1}\tilde{\mH}_{[i\:i+1]},~~{\rm where}\label{bondHam}\\
\f{\tilde{\mH}_{[i\:i+1]}}{\hbar}&=& 2J_x(\si_i^\dg\si_{i+1}+\si_{i+1}^\dg\si_i)+4J_z(\si_i^\dg\si_i\si_{i+1}^\dg\si_{i+1})\nn\\&&+\f{1}{2}(\delta\tilde{\om}_i\si_i^\dg\si_i+\delta\tilde{\om}_{i+1}\si_{i+1}^\dg\si_{i+1}),\label{bondHam1}
\eea
with $\delta\tilde{\om}_1=2(\om_1-\om_p),\delta\tilde{\om}_N=2(\om_N-\om_p)$ and $\delta\tilde{\om}_i=\om_i-\om_p$ for $i=2,3\dots N-1$. Due to such separability of the medium Hamiltonian, we can now write the commutation relation in Eq.~\ref{HEq1} as:
\bea
&&[\tilde{\boldsymbol{\sigma}}^{\otimes N},\tilde{\mH}_T]=[\tilde{\boldsymbol{\sigma}}^{\otimes N},\mH_{E}]+[\tilde{\boldsymbol{\sigma}}^{\otimes N},\tilde{\mH}_M]\nn\\
&&=[\tilde{\boldsymbol{\sigma}}^{\otimes N},\mH_{E}]+[\tilde{\boldsymbol{\sigma}}_1\otimes \tilde{\boldsymbol{\sigma}}_2,\tilde{\mH}_{[1\:2]}]\otimes\tilde{\boldsymbol{\sigma}}_3 \dots \otimes\tilde{\boldsymbol{\sigma}}_N\nn\\&&+\tilde{\boldsymbol{\sigma}}_1\otimes[\tilde{\boldsymbol{\sigma}}_2\otimes\tilde{\boldsymbol{\sigma}}_3,\tilde{\mH}_{[2\:3]}]\otimes\tilde{\boldsymbol{\sigma}}_4\dots\otimes\tilde{\boldsymbol{\sigma}}_N+\dots\nn\\
&&+\tilde{\boldsymbol{\sigma}}_1\otimes \tilde{\boldsymbol{\sigma}}_2\dots\otimes\tilde{\boldsymbol{\sigma}}_{N-2}\otimes[\tilde{\boldsymbol{\sigma}}_{N-1}\otimes \tilde{\boldsymbol{\sigma}}_N,\tilde{\mH}_{[N-1\:N]}].\label{comm1}
\eea

We can represent $\tilde{\mH}_{[i\:i+1]}$ as an MPO which helps to perform the commutation between the atomic operators and the nearest-neighbor bond Hamiltonian. Therefore, we express $\tilde{\mH}_{[i\:i+1]}$ as
\bea
\tilde{\mH}_{[i\:i+1]}&=&M^{[i]}M^{[i+1]},
\eea
where the matrix $M^{[i]}$ [$M^{[i+1]}$] acts on a local Hilbert space of $i$th [$(i+1)$th] atom, and the tensor product between elements of $M^{[i]}$ and $M^{[i+1]}$ generates the full Hilbert space of the bond Hamiltonian $\tilde{\mH}_{[i\:i+1]}$. We take matrix product between operator-valued matrices $M^{[j]}$'s which reproduces the bond Hamiltonian as sums of tensor products of operators represented by $2\times 2$ Pauli matrices in different local Hilbert spaces. We have the following operator-valued matrices for nearest-neighbor bond Hamiltonian:
\bea
M^{[i]}&=&\left[\ba{ccccc}\f{1}{2}\delta\tilde{\om}_i\si_i^\dg\si_i&2J_x\si_i^\dg&2J_x\si_i&4J_z\si_i^\dg\si_i&\mI_i\ea\right], \\
M^{[i+1]}&=&\left[\ba{ccccc}\mI_{i+1}&\si_{i+1}&\si_{i+1}^\dg&\si_{i+1}^\dg\si_{i+1}&\f{1}{2}\delta\tilde{\om}_{i+1}\si_{i+1}^\dg\si_{i+1}\ea\right]^T.\nn\\
\eea

For a homogeneous atomic medium with $\om_i=\om_a$ for all atoms, the commutation relation $[\tilde{\boldsymbol{\sigma}}^{\otimes N},\tilde{\mH}_M]$ can be evaluated easily. In this case, $\tilde{\mH}_{[i\:i+1]}$ is the same for all bulk bonds [for $i=2,3\dots (N-2)$], and we need to compute only a single commutator for all the bulk bonds. However, the above simplification does not work for a disordered medium, and we need to evaluate all the local commutators in Eq.~\ref{comm1} explicitly.

Next, we determine the contribution from the baths in the Heisenberg equations of the atomic operators. For this, we integrate out the photon baths at the left and right boundaries of the atomic medium by substituting the formal solutions (similar to Eqs.~\ref{HEsol1},\ref{HEsol2}) of the bath operators as we have done in Sec.~\ref{2atoms}. Thus, we find the following expression for the baths' contributions in the Heisenberg equations after performing an expectation in the coherent state of the incident light field: 
\bea
\la[\tilde{\boldsymbol{\sigma}}^{\otimes N},\mH_{E}]\ra=&&i\hbar \la \big((\boldsymbol{\Xi}_1-\boldsymbol{\Xi}_2)\otimes^{N}_{j=2}\tilde{\boldsymbol{\sigma}}_j\nn\\
&&+\otimes^{N-1}_{j=1}\tilde{\boldsymbol{\sigma}}_j\otimes (\boldsymbol{\Pi}_1-\boldsymbol{\Pi}_2)\big) \ra,\label{comm4}
\eea
where the components are:
\bea
\boldsymbol{\Xi}_1&=&\G_L\tilde{\si}_1^\dg[\tilde{\boldsymbol{\sigma}}_1,\tilde{\si}_1]-i\Om_L[\tilde{\boldsymbol{\sigma}}_1,\tilde{\si}_1],\nn\\
\boldsymbol{\Xi}_2&=&\G_L[\tilde{\boldsymbol{\sigma}}_1,\tilde{\si}_1^\dg]\tilde{\si}_1+i\Om_L[\tilde{\boldsymbol{\sigma}}_1,\tilde{\si}_1^\dg],\nn\\
\boldsymbol{\Pi}_1&=&\G_R\tilde{\si}_N^\dg[\tilde{\boldsymbol{\sigma}}_N,\tilde{\si}_N], \nn\\
\boldsymbol{\Pi}_2&=&\G_R[\tilde{\boldsymbol{\sigma}}_N,\tilde{\si}_N^\dg]\tilde{\si}_N\nn.
\eea

We take an expectation of the Heisenberg equation in Eq.~\ref{HEq1} in the coherent state of the incident light field:
\bea
\f{d\la \tilde{\boldsymbol{\sigma}}^{\otimes N} \ra}{dt} &=&-\f{i}{\hbar}\la [\tilde{\boldsymbol{\sigma}}^{\otimes N},\tilde{\mH}_T] \ra \nn \\
&=& -\f{i}{\hbar}\big(\la[\tilde{\boldsymbol{\sigma}}^{\otimes N},\mH_{E}]\ra + \la[\tilde{\boldsymbol{\sigma}}^{\otimes N},\tilde{\mH}_{M}]\ra\big).\label{comm5}
\eea
The different parts of the last line in Eq.~\ref{comm5} can be derived using  Eqs.~\ref{comm1}-\ref{comm4}. Here, $\la[\tilde{\boldsymbol{\sigma}}^{\otimes N},\tH_T]\ra$ is a tensor with indices as $\la[\tilde{\boldsymbol{\sigma}}^{\otimes N},\tH_T]\ra^{k_N}_{i_N,j_N}$ which we reshape into $\la[\tilde{\boldsymbol{\sigma}}^{\otimes N},\tH_T]\ra_{k_N,(i_N*j_N)}$. For a medium of $N$ atoms, $k_N=4^N$ and $i_N,j_N=2^N$. The reshaped tensor $\la[\tilde{\boldsymbol{\sigma}}^{\otimes N},\tH_T]\ra_{k_N,(i_N*j_N)}$ is a square matrix of dimension $4^N\times 4^N$. The basis for this $4^N\times 4^N$ square matrix is $4^N$ elements from $\la \tilde{\boldsymbol{\sigma}}^{\otimes N}\ra$. A row of the matrix $\la[\tilde{\boldsymbol{\sigma}}^{\otimes N},\tH_T]\ra_{k_N,(i_N*j_N)}$ corresponding to the element $\mI^{\otimes N}$ of $\la \tilde{\boldsymbol{\sigma}}^{\otimes N}\ra$ is null and we drop it. We also separate a column of the matrix $\la[\tilde{\boldsymbol{\sigma}}^{\otimes N},\tH_T]\ra_{k_N,(i_N*j_N)}$ corresponding to the element $\mI^{\otimes N}$ of $\la \tilde{\boldsymbol{\sigma}}^{\otimes N}\ra$, and write it as a column matrix $\boldsymbol{\Omega}_N$. For an incident light from the left of the atomic medium, there are two non-zero terms $\pm i\Omega_L$ in the column matrix $\boldsymbol{\Omega}_N$. After dropping a row and excluding a column from $\la[\tilde{\boldsymbol{\sigma}}^{\otimes N},\tH_T]\ra_{k_N,(i_N*j_N)}$, we now have a truncated square matrix of dimension $(4^N-1)\times (4^N-1)$ for $N$ atoms. Therefore, we find from Eq.~\ref{comm5}
\bea
\f{d\boldsymbol{\mathcal{S}}_N}{dt}=\boldsymbol{\mathcal{Z}}_N\boldsymbol{\mathcal{S}}_N+\boldsymbol{\Omega}_N,\label{sigma6}
\eea
where $\boldsymbol{\mathcal{S}}_N$ is a column matrix made of $(4^N-1)$ elements of $\la \tilde{\boldsymbol{\sigma}}^{\otimes N}\ra$ excluding the first element $\mI^{\otimes N}$. Here, $\boldsymbol{\mathcal{Z}}_N$ is the above truncated $(4^N-1)\times (4^N-1)$ square matrix obtained from $\la[\tilde{\boldsymbol{\sigma}}^{\otimes N},\tH_T]\ra_{4^N,4^N}$ after dropping a row and excluding a column. When we arrange $\la \sigma_1^{\dg}(t)\otimes \mI^{\otimes (N-1)}\ra e^{-i\om_p(t-t_0)}$ and $\la \sigma_1(t)\otimes \mI^{\otimes (N-1)}\ra e^{i\om_p(t-t_0)}$ as the first two elements of $\boldsymbol{\mathcal{S}}_N$, then we have $\boldsymbol{\Omega}_N=\begin{bmatrix}i\Omega_L &-i\Omega_L&0&0&\dots&0\end{bmatrix}^{T}$ of length $(4^N-1)$ for an incident light from the left of the atomic medium. 

As in the previous Sec.~\ref{2atoms}, we can calculate $\boldsymbol{\mathcal{S}}_N(t)$ from Eq.~\ref{sigma6} for an initial condition, e.g., $\boldsymbol{\mathcal{S}}_N(t_0)=0$ which denotes all atoms in the ground state at $t=t_0$ before shining a light on the atomic medium. Again, the long-time steady-state values of $\boldsymbol{\mathcal{S}}_N(t)$ are independent of the initial condition, and we evaluate $\boldsymbol{\mathcal{S}}_N(t =\infty)$ from
\bea
\boldsymbol{\mathcal{S}}_N(t=\infty)=-\boldsymbol{\mathcal{Z}}_N^{-1}\boldsymbol{\Omega}_N,\label{sigma7}
\eea
which requires an inversion of a square matrix of dimension $(4^N-1)\times (4^N-1)$ for $N$ atoms. The long-time steady-state transmission and reflection coefficients $\mathcal{T}_N(t=\infty)$ and $\mathcal{R}_N(t=\infty)$ of an incident laser light from the left of the atomic medium can be found from the following expressions:
\bea
\mathcal{T}_N(t=\infty)&=&\f{2\Gamma_R}{v_gI_{\rm in}}\mathcal{S}_{NN}(t =\infty),\label{transN} \\
\mathcal{R}_N(t =\infty)&=&1+\f{2\Omega_L}{v_gI_{\rm in}}{\rm Im}[\mathcal{S}_1(t= \infty)]\nn\\&&+\f{2\Gamma_L}{v_gI_{\rm in}} \mathcal{S}_{11}(t= \infty),\label{reflN}
\eea
where $\mathcal{S}_{NN}(t)=\langle \mI^{\otimes (N-1)}\otimes\sigma_N^{\dg}(t)\sigma_N(t) \rangle, \mathcal{S}_1(t)=\la \sigma_1(t) \otimes \mI^{\otimes (N-1)}\ra e^{i\om_p(t-t_0)}$ and $\mathcal{S}_{11}(t)= \la \sigma_{1}^{\dg}(t)\sigma_{1}(t)\otimes \mI^{\otimes (N-1)}\ra$ which are components of the column matrix $\boldsymbol{\mathcal{S}}_N$.

We here also calculate nonequilibrium properties of the driven-dissipative atomic medium. The local atomic excitation probability $\mathcal{S}_{ii}(t=\infty) \equiv \la \mI^{\otimes (i-1)}\otimes \si_i^\dg(t=\infty)\si_i(t=\infty)\otimes \mI^{\otimes (N-i)} \ra$ at the $i$th site is one such nonequilibrium feature of the medium which is already found in Eq.~\ref{sigma7}. The equal-time correlations between different atoms, for example, $\la\si_1^\dg(t)\si_1(t)\otimes \mI^{\otimes (N-2)}\otimes \si_N^\dg(t)\si_N(t)\ra$ -- an equal-time correlation between excited atoms at the boundaries of the medium, can also be extracted from Eq.~\ref{sigma7}. However, we need to go beyond the above calculation to derive a two-time correlation between atomic operators of the driven-dissipative medium. This can be formulated following Ref.~\cite{RoyPRA2017} where such two-time correlators are found for a single atom.

\begin{figure}
\includegraphics[width=0.99\linewidth]{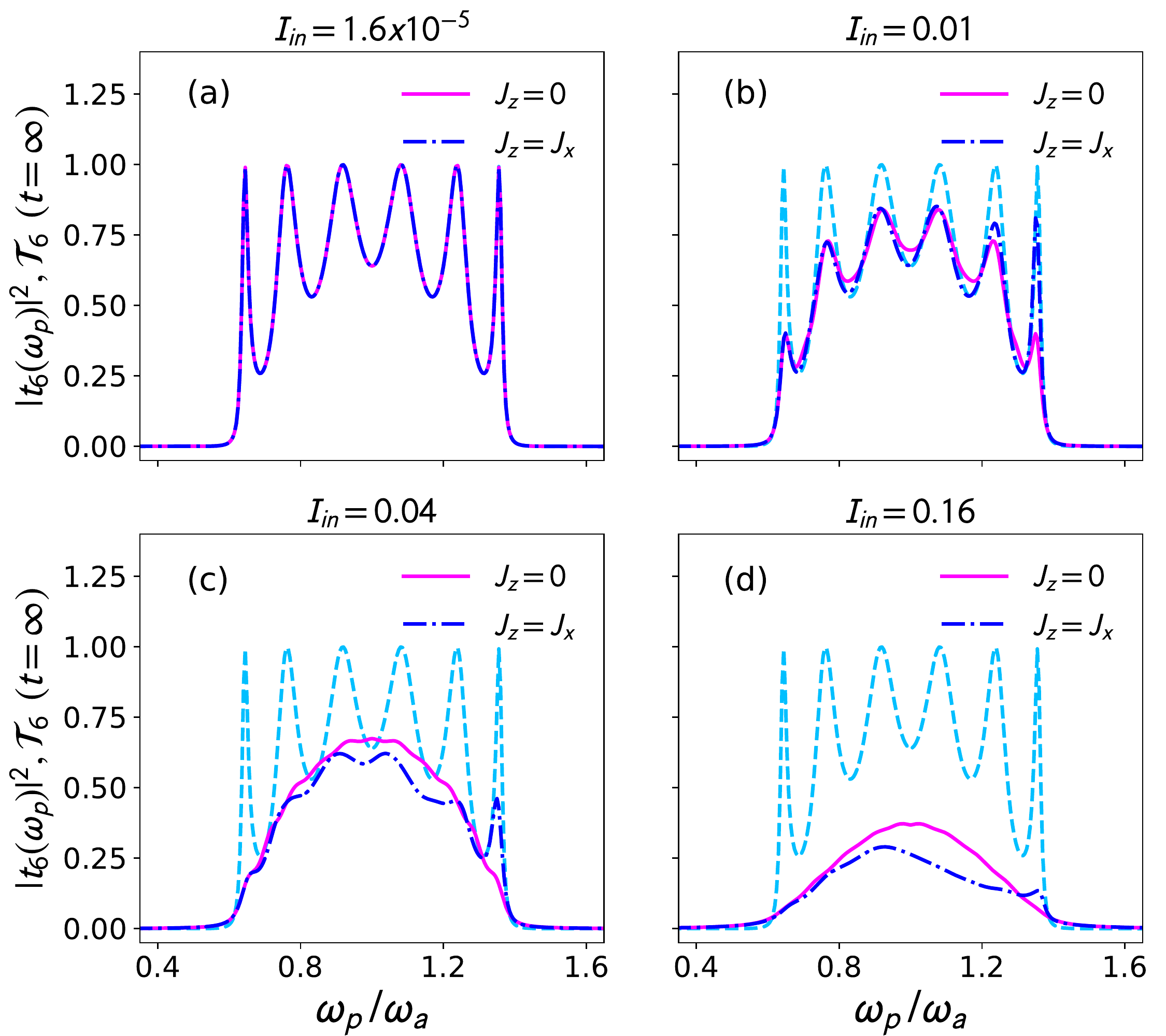}
\caption{Comparison between linear and nonlinear transmission lineshapes through an ordered atomic medium of six atoms modeled as an XX $(J_z=0)$ and an XXZ type interacting $(J_z \ne 0)$ spin chain. Different subplots are for different incident laser powers ($I_{\rm in}\propto E_p^2$). The single-photon (linear) transmission $|t_6(\om_p)|^2$ (dashed lines) is identical to the steady-state transmission coefficient $\mathcal{T}_6(t=\infty)$ for $I_{\rm in}=1.6 \times 10^{-5} \omega_a/v_g$. The parameters are $J_x=\Gamma_L=\Gamma_R=0.1\om_a$. The intensity $I_{\rm in}$ is in units of $\om_a/v_g$.} 
\label{NonSin}
\end{figure}

{\it Numerical results:} In the following, we present our numerical results for an ordered atomic chain with $\om_i=\om_a$ for $i=1,2\dots N$. We first discuss various features of the steady-state light transmission through the ordered medium and we later show the properties of the driven-dissipative medium. Our numerics for nonlinear light propagation following the QLE method in this section is currently limited to eight atoms due to the large size of $\boldsymbol{\mathcal{Z}}_N$ matrix in Eq.~\ref{sigma7}. However, we can study single-photon light propagation for a large number of atoms using the scattering theory in Appendix~\ref{App1}. Similar to the two atoms in Sec.~\ref{2atoms}, the single-photon transmission  $|t_N(\om_p)|^2$ through a medium of several atoms also has peaks around the resonant levels of the isolated medium modeled as an XX chain. It is depicted in all the subplots of Fig.~\ref{NonSin} by dashed lines for an ordered medium of six atoms. The number of peaks is equal to the number of atoms, and the frequency window ($\om_a-4J_{x}$ to $\om_a+4J_{x}$) of finite transmission in an ordered medium is determined by the coupling $J_{x}$. The width of the resonant peaks is controlled by the coupling strength $\Gamma_L$ and $\Gamma_R$, and the separation between the resonant peaks in the single-photon transmission decreases with increasing number of atoms. 

In Fig.~\ref{NonSin}, we also show how the single-photon transmission through six atoms changes with an increasing power of the incident light. The single-photon (linear) transmission $|t_6(\om_p)|^2$ from Appendix~\ref{App1} is identical to the steady-state transmission coefficient $\mathcal{T}_6(t=\infty)$ for a very low laser power (check Fig.~\ref{NonSin}(a)). We find that the sharp resonant peaks of the single-photon transmission lineshape start to disappear due to the saturation of $J_x$ coupling at a higher laser power which we have discussed for the two atoms in the previous section. The magnitude of nonlinear light transmission falls with an increasing power similar to the two atoms, and a broad and relatively smooth peak appears around atomic transition frequency $\om_a$ at a high laser power as presented in Figs.~\ref{NonSin}(c,d). The nonlinear $\mathcal{T}_6(t=\infty)$ is asymmetric around $\om_a$ for an interacting spin chain $(J_z \ne 0)$.

\begin{figure}
\includegraphics[width=1.0\linewidth]{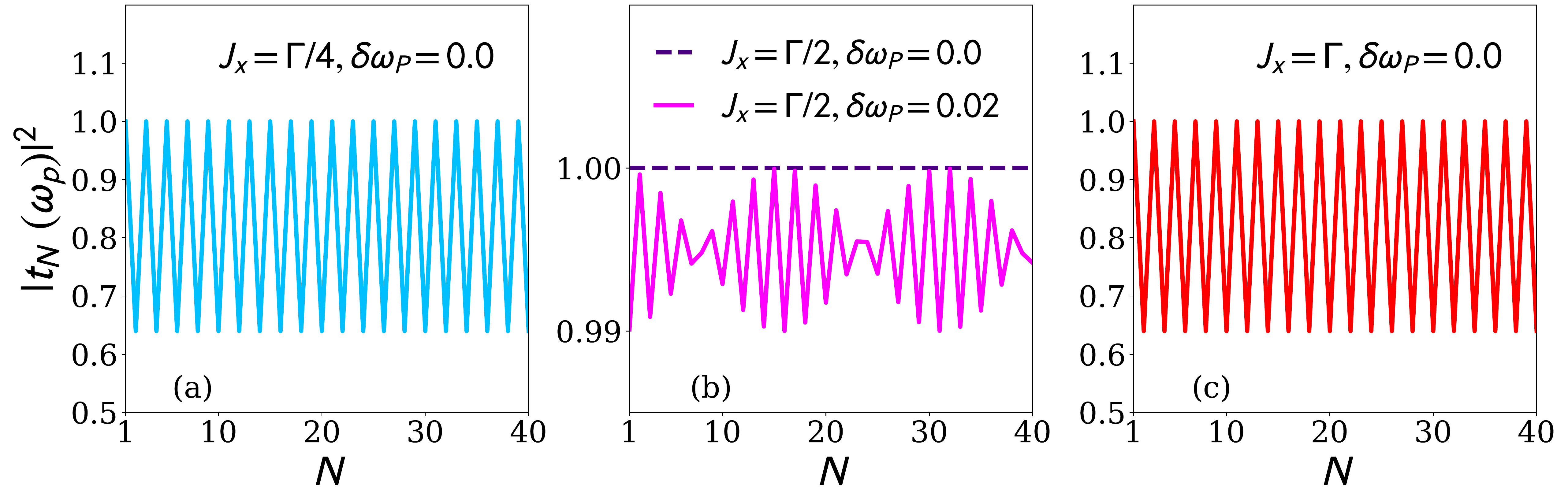}
\caption{Scaling of single-photon transmission $|t_N(\om_p)|^2$ with number of atoms $N$ of an ordered atomic medium modeled as an XX $(J_z= 0)$ or an XXZ like interacting $(J_z \ne 0)$ spin-1/2 chain. The detuning $\delta \om_p=(\om_a-\om_p)$ of the incident light from the atomic transition frequency $\om_a$ and the coupling $J_x$ are shown in the subplots. The parameters are $J_z=0.05\om_a, \Gamma_L=\Gamma_R=\Gamma=0.1\om_a$ and $\delta \om_p=0,0.02\omega_a$.} 
\label{SinPh}
\end{figure}

Next, we discuss the scaling of $|t_N(\om_p)|^2$ with $N$ in an ordered medium for a resonant (i.e., $\om_p=\om_a$) and a non-resonant (i.e., $\om_p \ne \om_a$) incident light. From Fig.~\ref{SinPh}, we find that the single-photon transmission is always ballistic in an ordered medium modeled as either an XX or an interacting spin chain. For a resonant light, the single-photon transmission coefficient $|t_N(\om_p)|^2$ is one when $2J_{x}=\Gamma_L=\Gamma_R=\Gamma$, and $|t_N(\om_p)|^2$ oscillates (without decay) with $N$ for a non-resonant light or when  $2J_{x} \ne \Gamma_L=\Gamma_R$. The latter oscillation for $2J_{x} \ne \Gamma_L=\Gamma_R$ is due to coherent scattering of light between two boundaries created by a mismatch between the $J_x$ coupling inside the medium and the tunnel coupling at the boundaries.

\begin{figure*}
\includegraphics[width=0.99\linewidth]{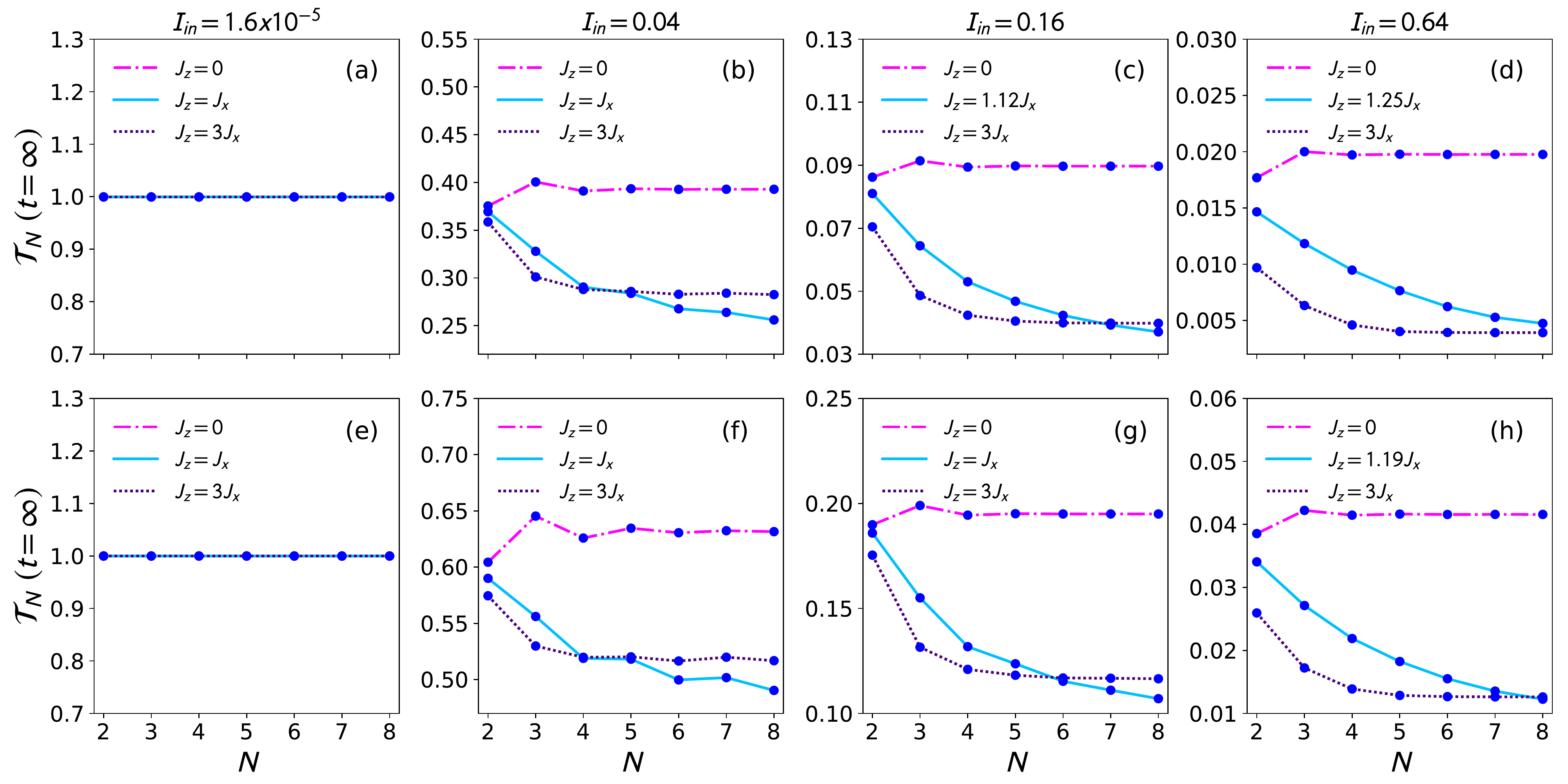}
\caption{Scaling of nonlinear transmission coefficient $\mathcal{T}_N(t=\infty)$ of a resonant monochromatic laser with the length $N$ of an ordered atomic media modeled as an XX $(J_z= 0)$ and an XXZ like interacting $(J_z \ne 0)$  spin-1/2 chain. Four columns show four different incident laser powers ($I_{\rm in}\propto E_p^2$). The parameters are $\om_p=\om_a, \Gamma_L=\Gamma_R=2J_x$, $J_x=0.025\om_a$ (1st row) and $0.05\om_a$ (2nd row). The intensity $I_{\rm in}$ is in units of $\om_a/v_g$.}
\label{Scaling}
\end{figure*}

We summarize the main features of the nonlinear transmission coefficient of a resonant laser through the ordered XX and interacting spin chains in Fig.~\ref{Scaling}. We find that the light transmission $\mathcal{T}_N(t=\infty)$ in an ordered XX spin chain ($J_z=0$) is independent of $N$ (ballistic) at any incident light power. It is depicted in the plots of Fig.~\ref{Scaling} for four different incident power (four columns of the figure) and two different $J_x$ (two rows). The oscillation of light transmission coefficient with $N$ at low power for $2J_{x} \ne \Gamma_L=\Gamma_R$ diminishes with an increasing light power both in the XX and interacting spin chains, and the nonlinear transmission lineshape changes monotonically with $N$.

\begin{figure}
\includegraphics[width=0.99\linewidth]{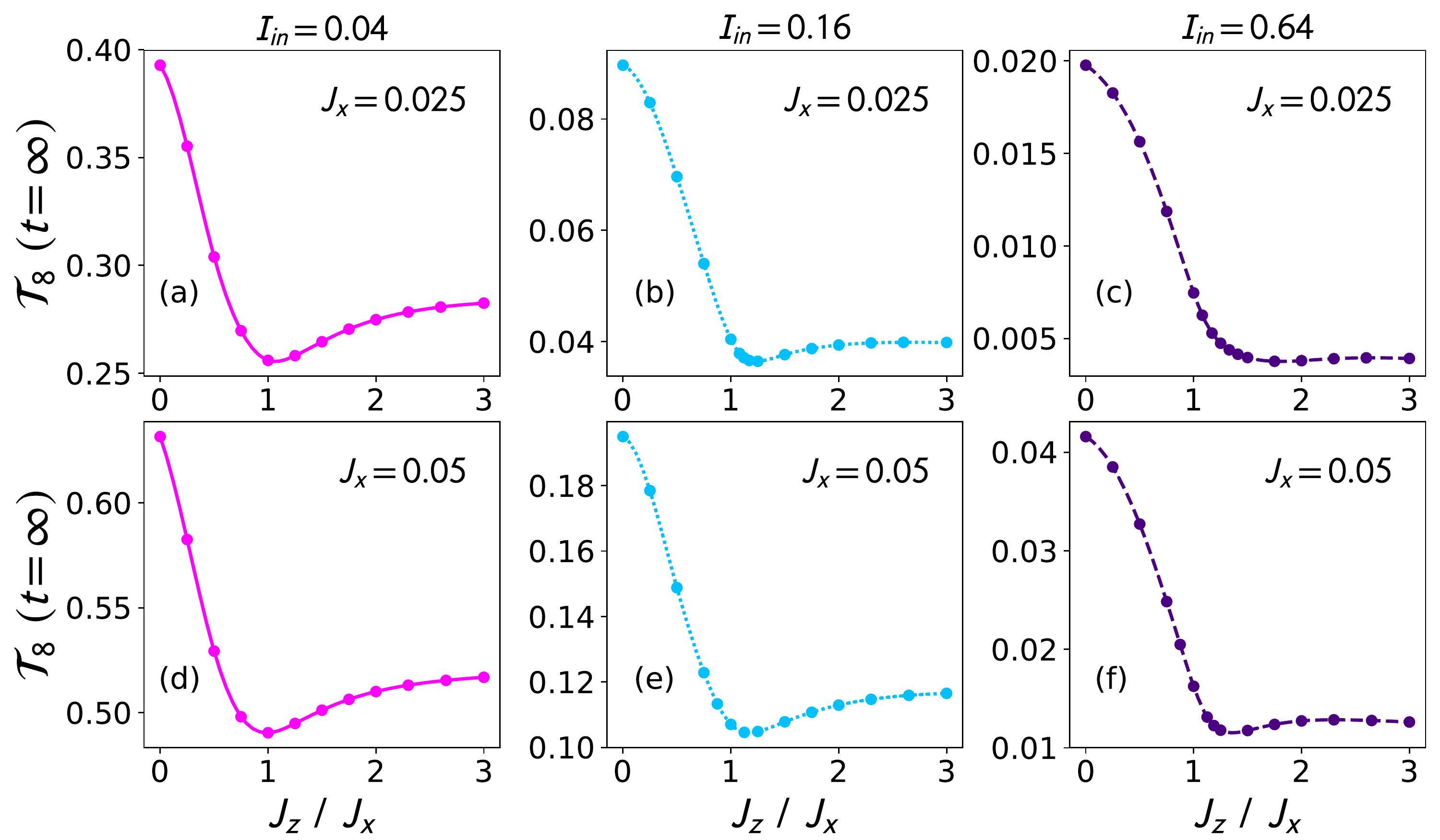}
\caption{Scaling of nonlinear transmission coefficient $\mathcal{T}_8(t=\infty) $ of a finite spin-1/2 chain of eight atoms with the ratio $J_z/J_x$ for three incident powers of a resonant laser. The lines joining the data points are some fits for a guide to the eye. The parameters are $\om_p=\om_a, \Gamma_L=\Gamma_R=2J_x$. The rate $J_x$ is in units of $\om_a$, and $I_{\rm in}$ is in units of $\om_a/v_g$.}
\label{trans8}
\end{figure}

The properties of $\mathcal{T}_N(t=\infty)$ of a resonant laser in an atomic medium modeled as an interacting spin chain ($J_z \ne 0$) are very interesting, and they depend on both the incident power and the value of $J_z$. We notice from Figs.~\ref{Scaling}(b,c,f,g) that the value of $\mathcal{T}_8(t=\infty)$ is a non-monotonic function of $J_z/J_x$ at a large light power. The nonlinear light transmission through a chain of eight atoms is higher for a small and a very large $J_z/J_x$ than an intermediate $J_z/J_x \approx 1$. Though this feature is not evident at an even higher power in Figs.~ \ref{Scaling}(d,h), but it seems to be the case in a longer chain (say of ten atoms) at this power when we follow the trend of $\mathcal{T}_N(t=\infty)$ in Figs.~ \ref{Scaling}(d,h). These features are sketched in Fig.~\ref{trans8}.

\begin{figure}
\includegraphics[width=0.99\linewidth]{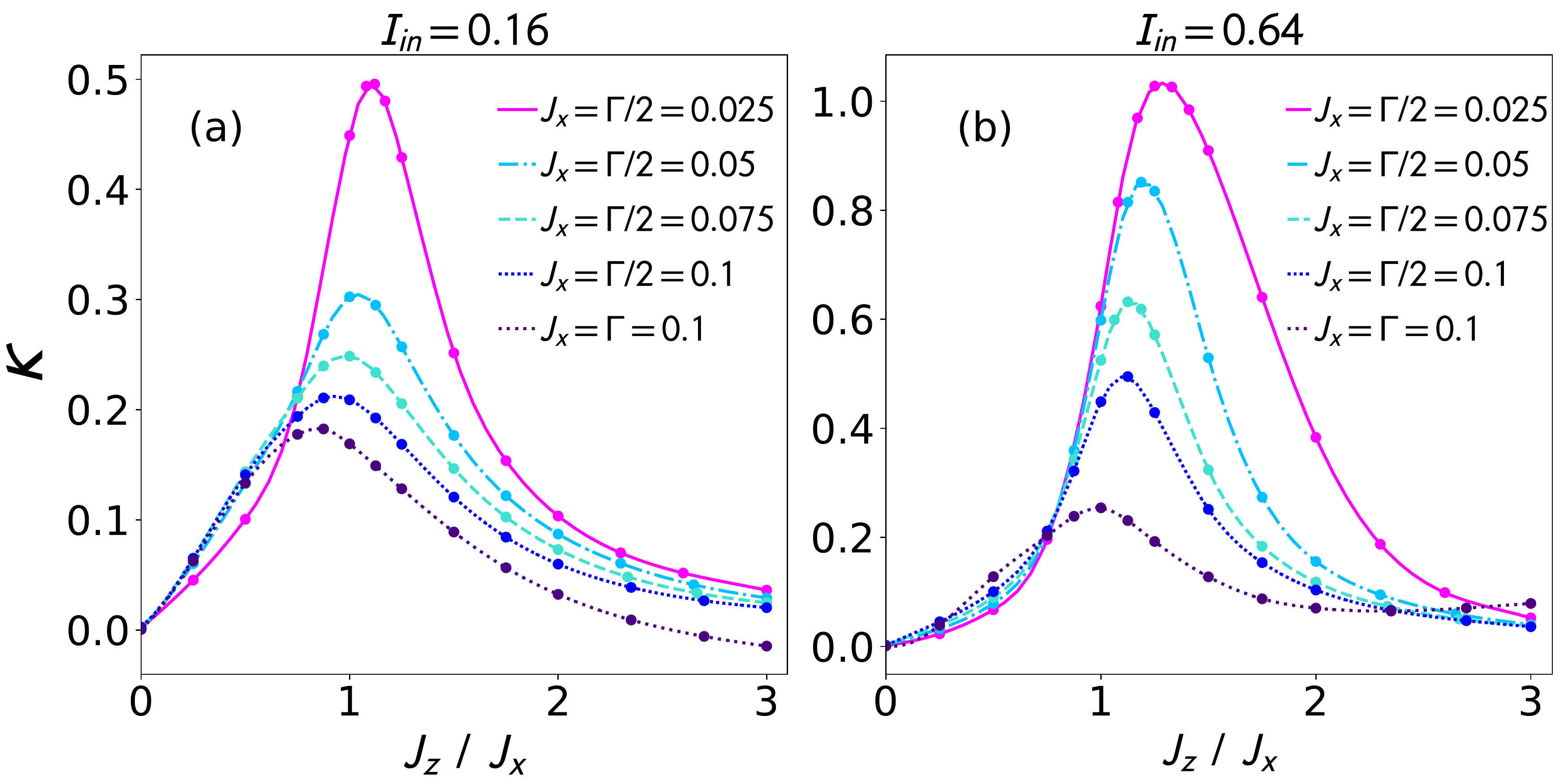}
\caption{Dependence of the scaling exponent $\kappa$ in $\mathcal{T}_N(t=\infty) \propto N^{-\kappa}$ on the ratio $J_z/J_x$ of an interacting atomic medium for two incident powers of a resonant laser. The lines joining the data points are some fits for a guide to the eye. The parameters are $\om_p=\om_a, \Gamma_L=\Gamma_R=\Gamma$. The rates $J_x$ and $\Gamma$ are in units of $\om_a$, and $I_{\rm in}$ is in units of $\om_a/v_g$.}
\label{kappa}
\end{figure}

Like the XX spin chain,  we find that the nonlinear $\mathcal{T}_N(t=\infty)$ is independent of $N$ (ballistic) for $J_z/J_x \gg1 $. It is depicted in Figs.~\ref{Scaling}(b,c,d,f,g,h) for $J_z/J_x=3$ with two different values of $J_x$. The scaling of $\mathcal{T}_N(t=\infty)$ with $N$  at some intermediate values of $J_z$ when $J_z/J_x \approx 1$ is a bit non-trivial, and it is difficult to predict the asymptotic scaling from our numerics with short chains. Nevertheless, we find a clear $N$-dependence of $\mathcal{T}_N(t=\infty)$ in finite-size interacting spin chains at a high light power. We extract the exponent $\kappa$ in $\mathcal{T}_N(t=\infty) \propto N^{-\kappa}$, and present $\kappa$ with $J_z/J_x$ in Fig.~\ref{kappa} for two different laser powers. The critical value of  $J_z/J_x$ (called $\Delta_c$) where $\kappa$ is maximum, seems to depend on the incident light power and $J_x$. The value of $\Delta_c$ increases with a higher power or a smaller $J_x$ which are shown in Fig.~\ref{kappa}. We add $\mathcal{T}_N(t=\infty)$ with $N$ at $\Delta_c$ in Figs.~\ref{Scaling}(b,c,d,f,g,h). We also find the value of $\kappa$ at $\Delta_c$ increases with  smaller values of $J_x$ and higher values of $\Gamma_L,\Gamma_R$. The Fig.~\ref{kappa} shows that for a fixed value of $J_x$, the maximum value of $\kappa$ is higher for $\Gamma_L,\Gamma_R=2J_x$ than that for $\Gamma_L,\Gamma_R=J_x$. The above-discussed features of nonlinear light transmission of a resonant laser seem to remain the same for a small detuning of the incident light $(\om_p \ne \om_a)$.

Our studied model of an interacting spin chain here is a bit similar to the Heisenberg XXZ model in external magnetic fields. One distinction between our interacting spin chain and the XXZ model in external magnetic fields is that we have dropped a rescaling of the local fields $\om_i$ when we transform the Pauli matrices by the ladder operators in the Heisenberg XXZ chain to write the Hamiltonian in Eq.~\ref{Hams}. The dropping of such rescaling in the local fields or the transition frequencies helps us to find the length dependence of nonlinear transmission coefficient for a resonant light at a fixed frequency $\om_p$. The isolated XXZ spin-1/2 chain in the absence of external fields $(\om_i=0)$ shows a ground-state (equilibrium) phase transition between disordered paramagnet for $J_z/J_x<1$ and ordered antiferromagnet for $J_z/J_x>1$ when $J_x,J_z>0$. Such ground-state phase transition survives in the presence of a uniform external field, but the critical value of $J_z/J_x$ for the transition depends on the external field. 

\begin{figure*}
\includegraphics[width=0.99\linewidth]{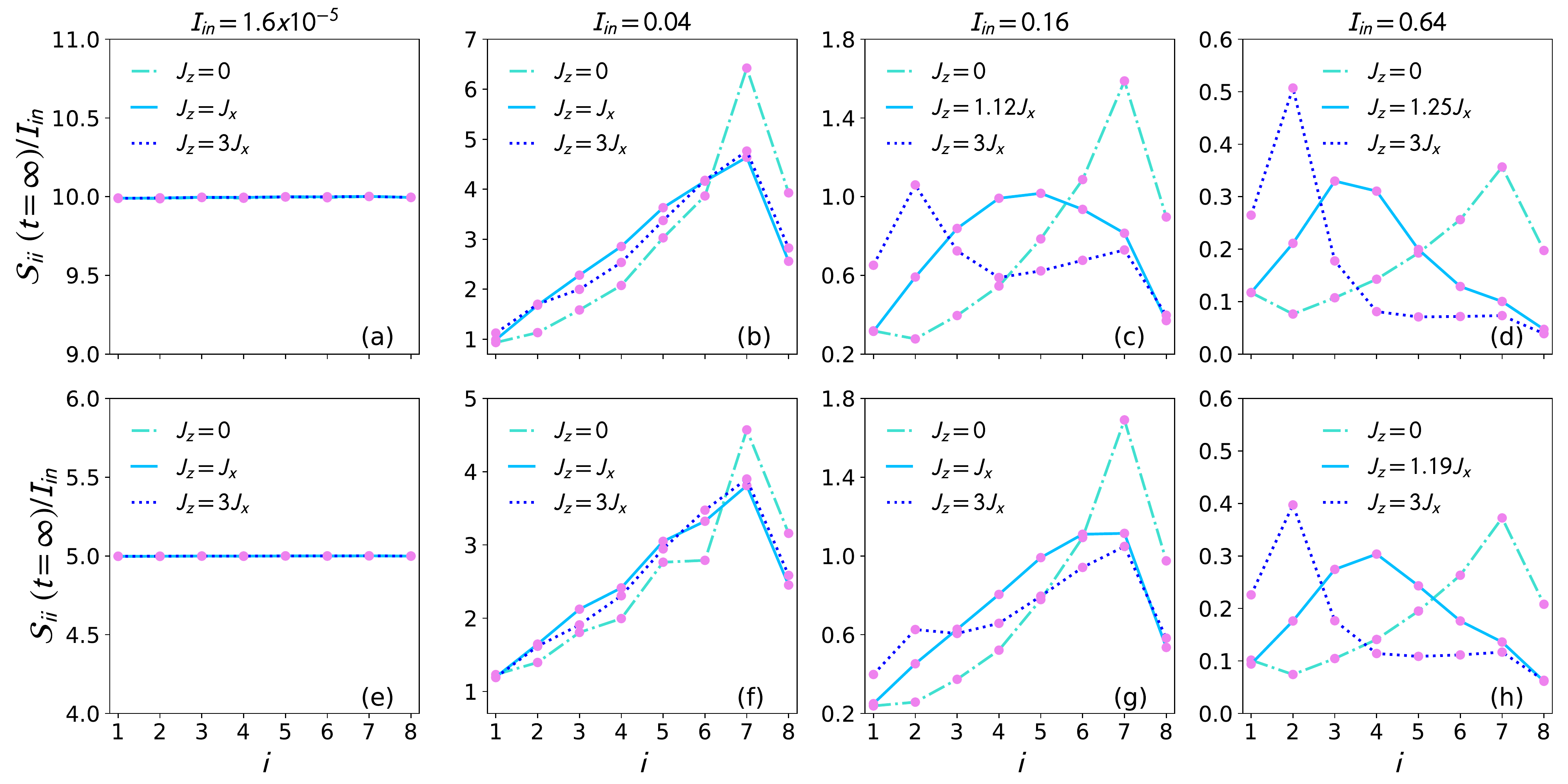}
\caption{Scaled atomic excitation $\mathcal{S}_{ii}(t=\infty)/I_{\rm in}$ of an ordered atomic medium of eight atoms $(N=8)$ modeled as an XX $(J_z=0)$ and an XXZ like interacting $(J_z \ne 0)$ spin chain. Different columns are for different power of the incident laser ($I_{\rm in}\propto E_p^2$). The parameters are $\om_p=\om_a, \Gamma_L=\Gamma_R=2J_x$, $J_x=0.025\om_a$ (1st row) and $0.05\om_a$ (2nd row). The intensity $I_{\rm in}$ is in units of $\om_a/v_g$.}
\label{LocalEx}
\end{figure*}

The features of our driven-dissipative atomic medium are expected to be very different from that of the isolated medium. First, the properties of the nonequilibrium medium are determined by the scattering states of the open medium near the frequency $\om_p$ of the laser instead of the ground state of the isolated medium. With a stronger driving (higher $I_{\rm in}$) by the laser, more and more scattering states around $\om_p$ contribute in nonlinear light transmission. For a resonant light on an ordered medium, both the transition frequencies of the 2LAs $(\om_i=\om_a$ for $i=1,2 \dots N)$ and the frequency $\om_p$ (or $v_gk$) of the incident light disappear from the dynamical equations of the nonequilibrium medium which is evident in the matrix $\boldsymbol{\mathcal{Z}}_{s}$ for the single-photon transport and the Eqs.~\ref{s8}-\ref{s16} and Eq.~\ref{bondHam1} for nonlinear transport. Therefore, the nature of light transmission mainly depends on the coupling $J_x$ and $J_z$ of a long medium where the boundary coupling $\Gamma_L$ and $\Gamma_R$ play a less significant role.
In the absence of external fields, an isolated XXZ spin-1/2 chain has two trivial limits to the XX spin chain and the Ising spin chain respectively for $J_z=0$ and $J_x=0$. The XX spin chain can be mapped to a tight-binding chain of free fermions. We find that the nonlinear transmission coefficient in the XX spin chain falls with an increasing light power, but it is independent of the chain length. On the contrary, an Ising spin chain does not contribute to light transmission at any power. However, it can transmit light in the presence of a weak perturbation from the $J_x$ coupling, and we find a ballistic transport of a resonant light when $J_z/J_x \gg 1$. At an intermediate value of $J_z$ when $J_z$ and $J_x$ are comparable, the propagating photons in nonlinear light transmission experience maximum scattering, and the cusp in $\kappa$ vs. $J_z/J_x$ plots around $J_z/J_x \approx 1$ in Fig.~\ref{kappa} for a finite-size chain may be due to it.  

In Fig.~\ref{LocalEx}, we show the scaled atomic excitation $\mathcal{S}_{ii}(t=\infty)/I_{\rm in}$ of the $i$th atom of an ordered atomic medium modeled as an XX $(J_z=0)$ and an interacting $(J_z \ne 0)$ spin chain driven by a resonant laser of various power. We find while the profile of $\mathcal{S}_{ii}(t=\infty)/I_{\rm in}$ is similar in an XX and an interacting spin chain at a low light power (see Figs.~\ref{LocalEx}(a,e)), it can be notably different at a high light power in the absence and presence of interaction especially as shown in Figs.~\ref{LocalEx}(c,d,g,h). For a weak light power (linear regime), we find $\mathcal{S}_{ii}(t=\infty)/I_{\rm in}$ oscillates with the position of the atom when $\Gamma_L,\Gamma_R \ne 2J_x$ (not shown) and it remains flat when $\Gamma_L,\Gamma_R = 2J_x$ (see Figs.~\ref{LocalEx}(a,e)). This is a single-photon regime, and the interaction does not play a role here. All the atoms are equally excited by a propagating photon.

For a stronger driving by the laser field as in Figs.~\ref{LocalEx}(b,f), the value of atomic excitation increases from the left of the medium, where the laser is shined, to the right with an abrupt fall at the end. The atoms are more excited on the opposite side of the incoming light, and the maximum value of excitation is lower than the single-photon regime. The lower excitation at the left of the medium is probably due to the saturation of $J_x$ coupling at a higher power of the light. Such saturation also reduces the effective value of $2J_x$ between the last two atoms and it becomes smaller than $\Gamma_R$. It leads to leaking of excitation at the right boundary as the last atom is relatively strongly coupled to the right bath than the inner atom. The interacting chains seem to have a more linearized excitation profile than the noninteracting XX chains at this light power.

The shape of $\mathcal{S}_{ii}(t=\infty)/I_{\rm in}$ in a noninteracting chain remains almost the same with a further increase in the incident light power. However, the magnitude of the excitation falls with a higher power. These features are depicted in Figs.~\ref{LocalEx}(c,d,g,h). On the other hand, a strong interaction changes the shape of $\mathcal{S}_{ii}(t=\infty)/I_{\rm in}$ drastically in an interacting chain at a higher power. For example, we show in Figs.~\ref{LocalEx}(d,h) that the atomic excitation is mostly localized near the left boundary and is almost flat on the rest of the medium. It is because a strong repulsive interaction between atomic excitation prevents light from propagating through the medium. The shapes of the $\mathcal{S}_{ii}(t=\infty)/I_{\rm in}$ at an intermediate interaction $(J_z \approx J_x)$ in Figs.~\ref{LocalEx}(c,d,g,h) at two different power also follow the above argument. 

Finally, we discuss some crucial features which are intrinsic to the experimental realizations of such atomic media and are not included in our simple model in Eq.~\ref{Ham}. For example, most experiments with real or artificial atoms in 1D involve various dissipation and decoherence mechanisms such as pure dephasing, nonradiative decay and spontaneous emission into photon modes outside of the 1D continuum \cite{RoyRMP2017}. While pure-dephasing dominates in superconducting qubits \cite{vanLoo2013, Fitzpatrick2017}, nonradiative decay is common in most semiconductor quantum dots and molecules. The spontaneous emission into photon modes outside of the 1D continuum accounts for the typical loss of photons in experiments with Rydberg atoms driven by tightly focused laser lights \cite{Peyronel12, Firstenberg13}, quantum dots coupled to nanowire waveguides and quantum dots coupled to line-defects in photonic crystals \cite{Lodahl2015}. In Appendix~\ref{App2}, we incorporate some of these dissipation and decoherence mechanisms in our model in Eq.~\ref{Ham}, and study their effects on light transmission through the atomic media. 

We have considered only nearest-neighbor (short-range) interaction between atoms in our model for the atomic medium in Eq.~\ref{Hams}. However, it is not a good approximation for many experimental systems especially with Rydberg atoms \cite{Peyronel12, Firstenberg13} and polar molecules \cite{Yan2013} where long-range interactions between atoms such as dipole-dipole interactions are necessary to include. We show how to apply our generalized QLE method in the presence long-range interactions in Appendix~\ref{App3} and discuss some consequences of the long-range interactions in linear and nonlinear light transmission through the atomic media.

\section{Conclusions}
\label{conl}
In this paper, we have extended the application of QLE approach to a 1D interacting quantum optical medium. We investigate nonequilibrium dynamics of an interacting quantum medium driven by light fields and properties of the scattered light. In the recent years, there are several studies along this direction using different approaches such as the scattering theory \cite{RoyNat2013, Fang2015}, the input-output theory \cite{Lalumiere2013, Caneva2015}, the time-dependent Luttinger-liquid theory \cite{Otterbach2013}, the time-evolution of wave-function using matrix product states \cite{Manzoni2017}. Many of these studies are limited to only a few photons or a few atoms. Here, we can examine quantum optical nonlinearity  of the atomic medium  at any power of the incident laser and relatively large number ($N=8$) of atoms. One advantage of the QLE approach over the Lindblad master equation formalism is that the former method in contrast to the latter formalism is in principle valid for an arbitrary bath-medium coupling strength \cite{Breuer06, Purkayastha2016}. We have shown in Fig.~\ref{kappa} that the value of the scaling exponent $\kappa$ strongly depends on the ratio of the bath-medium coupling and the coupling between atoms in the medium. Therefore, the nature of the light transmission in the open quantum systems can change with that ratio, and this proves the usefulness of the QLE approach to explore nonequilibrium transport in such systems for general baths. 

Light propagation in a nonlinear optical medium is commonly studied using a nonlinear wave equation, and the properties of the medium are expressed through linear and nonlinear optical susceptibilities and refractive indices. One primary goal of our present work is to illuminate the microscopic origin of such optical susceptibilities and refractive indices in some simple models of a quantum optical medium of interacting atoms. The scaling of transmission coefficient with the number of atoms and the power of incident light for different media elucidates how macroscopic susceptibilities and refractive indices emerge from the microscopic interactions between individual atoms. 

While we can explore both the transient and steady-state dynamics of an interacting open quantum system, our current numerical results are limited to a 1D lattice model of eight sites. This limitation is at present the central challenge of our generalized QLE approach. Due to such restriction on the size of the medium, we can not determine the asymptotic scaling of the transmission coefficient with the system size. Therefore, we are not able to study any thermodynamic properties of the nonequilibrium quantum medium such as nonequilibrium phase transition. We emphasize here that the MPO representation helps us to obtain the matrix equation  in Eq.~\ref{sigma7} in a simple, compact form after integrating out the photon fields. However, the limitation on available system size for steady-state dynamics stems from the constraint in numerics to invert a $(4^N-1)\times(4^N-1)$ square matrix $\boldsymbol{\mathcal{Z}}_N$ in Eq.~\ref{sigma7}. Soon, we wish to test more sophisticated methods such as the Lanczos algorithm to increase the accessible size of $\boldsymbol{\mathcal{Z}}_N$. It might also be possible to extend some of the hypotheses in Refs.~\cite{ProzenJSM2009, ProsenPRL2011a, ProsenPRL2011b} using MPO ansatz to our generalized QLE approach. 

In our model in Eq.~\ref{Ham}, we connect the photon baths at the boundaries of the atomic medium, and an incident light can pass through the medium after interacting with all the atoms. We also assume any photon mediated interaction between atoms to be included in the $J_x$ and $J_z$ couplings. Nevertheless, there are other physical situations where light can be side-coupled to the atoms, and an incident light can propagate through a medium without interacting with some or all atoms \cite{Peyronel12, Firstenberg13}. Shortly, we wish to apply the QLE approach for such side-coupled atom-photon interaction which is also common in many experimental set-ups. We have tried to explore the role of disorder or randomness in the medium on the nonlinear light propagation. However, our results are inconclusive due to the constraint on chain length; so we do not present them here. Disordered medium would be an exciting direction of study when a longer chain length becomes accessible in the numerics. Finally, it would be enlightening to examine light propagation through an atomic medium by going beyond the Markov, and rotating-wave approximation assumed in writing the model Hamiltonian in our present study.

\section*{Acknowledgments}
We thank A. Dhar, R. Moessner and R. Singh for discussion. DR acknowledges funding from the Department of Science and Technology, India via the Ramanujan Fellowship.

\appendix
\setcounter{figure}{0}
\renewcommand\thefigure{A\arabic{figure}}

\section{Single-photon transport}
\label{App1}
The propagation of a single-photon through an atomic chain of $N$ 2LAs can be calculated using the Lippmann-Schwinger scattering theory \cite{Lippmann1950, RoyNat2013}. For a linear energy-momentum dispersion  of photons, we get the following real-space description of the Hamiltonian in Eq.~\ref{Ham} after replacing the photon operators in momentum space by their Fourier transform to real-space operators:\cite{RoyRMP2017}
\bea
\f{\mathcal{H}_R}{\hbar}&=&\f{\mathcal{H}_M}{\hbar}-iv_g\int dx \big(a_x^\dg\f{\partial}{\partial x}a_x+b_x^\dg\f{\partial}{\partial x}b_x\big)\nn\\
&&+(\sqrt{2v_g\Gamma_L}\:\si_1^\dg a_0 +\sqrt{2v_g\Gamma_R}\:\si_N^\dg b_0+h.c.).\label{HamA}
\eea
Let us consider a single-photon incident light from the left of the atomic medium. The wavefunction of a single incident photon with all the atoms in the ground state is
\bea
|\psi\ra_{\rm in}=\f{1}{\sqrt{2\pi}}\int dx\:e^{ikx}a_x^\dg|\varphi\ra \otimes|g,g,g\dots g\ra,
\eea
where $\omega_k~(\omega_k=v_gk)$ and $k$ are respectively the frequency and the wave vector of the incident photon, and  $|\varphi\ra$ denotes vacuum of the light fields. A single photon can excite the $i$th 2LA to excited state $|e\ra_i$. Therefore, we take an ansatz for the scattered wavefunction $|\psi\ra_{\rm s}$ as
\bea
|\psi\ra_{\rm s}&=&\f{1}{\sqrt{2\pi}}\int dx\:\{\delta(x)[e_k^1|\varphi \ra \otimes |e,g,g\dots g\ra \nn\\&& +e_k^2|\varphi\ra \otimes |g,e,g\dots g\ra+ \dots +e_k^N|\varphi\ra \otimes |g,g,g\dots e\ra]\nn\\
&&+(\phi_k^a(x)a_x^\dg+\phi_k^b(x)b_x^\dg)|\varphi\ra \otimes |g,g,g\dots g\ra\},
\eea
where $e_k^i$ is the amplitude for $i$th atom in the excited state $|e\ra_i$, and $\phi_k^a(x)$ and $\phi_k^b(x)$ are the amplitudes for a scattered photon in the left and right side of the atomic medium respectively.

We can find these amplitudes using the Schr{\"o}dinger equation $\mathcal{H}_{R}|\psi\ra_s=\hbar v_gk|\psi\ra_s$ which results in $N$ linear coupled equations plus two linear inhomogeneous differential equations. The differential equations are
\bea
iv_g\f{\partial}{\partial x}\phi_k^a(x)+v_gk\phi_k^a(x)=\sqrt{2v_g\Gamma_L}\delta(x)e_k^1, \label{SPhAm1}\\
iv_g\f{\partial}{\partial x}\phi_k^b(x)+v_gk\phi_k^b(x)=\sqrt{2v_g\Gamma_R}\delta(x)e_k^N.\label{SPhAm2}
\eea
The Eqs.~\ref{SPhAm1},\ref{SPhAm2} show that the single-photon amplitudes have a discontinuity at $x=0$: $\phi_k^a(0+)-\phi_k^a(0-)=-i\sqrt{2\Gamma_L/v_g}\:e_k^1$ and $\phi_k^b(0+)-\phi_k^b(0-)=-i\sqrt{2\G_R/v_g}\:e_k^N$. For an incident single-photon from the left of the medium: $\phi_k^a(0-)=1$ and $\phi_k^b(0-)=0$. Thus, we write
\bea
\phi_k^a(0)&=&\f{1}{2}(\phi_k^a(0+)+\phi_k^a(0-))=1-i\sqrt{\f{\Gamma_L}{2v_g}}e_k^1,\nn\\
\phi_k^b(0)&=&\f{1}{2}(\phi_k^b(0+)+\phi_k^b(0-))=-i\sqrt{\f{\Gamma_R}{2v_g}}e_k^N,\nn
\eea
by using regularization of the amplitudes across $x=0$.
We substitute the above $\phi_k^a(0)$ and $\phi_k^b(0)$ in $N$ linear coupled equations, obtained from the Schr{\"o}dinger equation, for the amplitudes $e_k^i$ with $i=1,2 \dots N$, and we write these equations in a compact form as $\boldsymbol{\mathcal{Z}}_{s}\bold{e}_k=\boldsymbol{\Omega}_{s}$ where $\boldsymbol{\mathcal{Z}}_{s}$ is a tridiagonal square matrix with elements $[\boldsymbol{\mathcal{Z}}_{s}]_{mm}=\om_m-v_gk-i\Gamma_{L}\delta_{m,1}-i\Gamma_{R}\delta_{m,N}$ and $[\boldsymbol{\mathcal{Z}}_{s}]_{mn}=2J_x(\delta_{m+1,n}+\delta_{m,n+1})$ for $m \ne n =1,2\dots N$, and 
\bea
\bold{e}_k&=&\begin{bmatrix} e_k^1&e_k^2&e_k^3&\dots&e_k^N\end{bmatrix}^{T},\label{emat} \\
\boldsymbol{\Omega}_{s}&=&\begin{bmatrix}-\sqrt{2v_g\Gamma_L}&0&0&\dots&0\end{bmatrix}^{T}.\label{omat}
\eea
The matrix $\boldsymbol{\mathcal{Z}}_{s}$ does not contain $J_z$ as the contribution from the z-component of the interaction  in the Hamiltonian in Eq.~\ref{Ham} or \ref{HamA} is zero for a single-photon incident state with all atoms in the ground state. We get the amplitudes of excited atom: $ \bold{e}_k=\boldsymbol{\mathcal{Z}}^{-1}_{s}\boldsymbol{\Omega}_{s}$. The single-photon transmission and reflection amplitudes $t_N(\omega_k),r_N(\omega_k)$ are obtained by using the solution of $e_k^1$ and $e_k^N$. They are
\bea
t_N(\omega_k)&=&-i\sqrt{2\G_R/v_g}\:e_k^N, \label{stra}\\
r_N(\omega_k)&=&1-i\sqrt{2\Gamma_L/v_g}\:e_k^1. \label{sref}
\eea
\section{Inclusion of losses from atoms}\label{App2}
In the main text, we have only considered dissipation and decoherence of the atomic medium due to its coupling to the light fields at the boundaries. However, light-propagation in an atomic medium can induce other types of dissipation and decoherence of real and artificial atoms in the bulk and edges of the medium. We here incorporate two such mechanisms namely nonradiative decay and pure dephasing in our microscopic analysis of light propagation. The total Hamiltonian of the light-matter system including the above losses is $\mathcal{H}_{\rm loss}=\mathcal{H}_T+\mH_{D}$ where the Hamiltonian $\mathcal{H}_T$ is given in Eq.~\ref{Ham} and
\bea
\f{\mH_{D}}{\hbar}&=&\sum_{j=1}^N\int_{-\infty}^{\infty}dk \big[v_gk(c_{jk}^{\dg}c_{jk}+d_{jk}^{\dg}d_{jk})\nn \\&&+\lambda (c_{jk}+c_{jk}^{\dg})\sigma_j^{\dg}\sigma_j+\gamma(d_{jk}^{\dg}\sigma_j+\sigma_j^{\dg}d_{jk})\big].
\eea
Here, the operators $c_{jk}^{\dg}, d_{jk}^{\dg}$ create excitations with wave vector $k$, respectively, related to pure dephasing and nonradiative decay of $j$th atom \cite{Koshino12,RoyPRA2017}. The rate of pure dephasing and nonradiative decay are respectively $\lambda$ and $\gamma$ which we take to be same for all atoms. The spontaneous decay rate of an excited atom to non-guided (outside the left and right photon baths) photon modes can also be included within the nonradiative decay rate $\gamma$. 

For an atomic medium of $N$ atoms, the time evolution of operators $c_{jk}(t)$ and $d_{jk}(t)$ with initial conditions $c_{jk}(t_0)$ and $d_{jk}(t_0)$ at an initial time $t=t_0$ are given by, 
\bea
c_{jk}(t)&=&\mG_k(t-t_0)c_{jk}(t_0)\nn\\&&-i\lambda\int_{t_0}^{t}dt'\mG_k(t-t')\si_j^{\dg}(t')\si_j(t'),\label{cop}\\
d_{jk}(t)&=&\mG_k(t-t_0)d_{jk}(t_0)\nn\\&&-i\g\int_{t_0}^{t}dt'\mG_k(t-t')\si_j(t'),\label{dop}
\eea
for $j=1,2,\dots N$, where again $\mG_k(\tau)=e^{-iv_gk\tau}$. As earlier, for an incident laser beam in a coherent state $|E_p,\om_p\ra$ from the left of the atomic medium, we have,
\bea
c_{jk}(t_0)|E_p,\om_p\ra=0,~ d_{jk}(t_0)|E_p,\om_p\ra=0.
\eea
In the presence of pure dephasing and nonradiative decay, the Eq.~\ref{comm5} for the atomic operators would have some extra contributions from $\mH_{D}$. Thus, we have
\bea
\f{d\la \tilde{\boldsymbol{\sigma}}^{\otimes N} \ra}{dt}&=&-\f{i}{\hbar}\big(\la[\tilde{\boldsymbol{\sigma}}^{\otimes N},\mH_{E}]\ra + \la[\tilde{\boldsymbol{\sigma}}^{\otimes N},\tilde{\mH}_{M}]\ra\nn\\&&+\la[\tilde{\boldsymbol{\si}}^{\otimes N},\mH_{D}]\ra\big).\label{comm6}
\eea
The first two terms in Eq.~\ref{comm6} can be found using Eqs.~\ref{comm1}-\ref{comm4}. To calculate $\la[\tilde{\boldsymbol{\si}}^{\otimes N},\mH_{D}]\ra$, we integrate out the excitation fields related to pure dephasing and nonradiative decay by substituting the formal solutions of $c_{jk}(t)$ and $d_{jk}(t)$ in Eqs.~\ref{cop},\ref{dop} as in Sec.~\ref{Natoms}. Therefore, we have
\bea
&&\la[\tilde{\boldsymbol{\si}}^{\otimes N},\mH_{D}]\ra/(i\hbar)\\&&=\la\G_{\lambda}\Big(\big[\tilde{\si}_1^{\dg}\tilde{\si}_1,[\tilde{\boldsymbol{\si}}_1,\tilde{\si}_1^{\dg}\tilde{\si}_1]\big]\otimes
\tilde{\boldsymbol{\si}}_2\otimes\dots\otimes\tilde{\boldsymbol{\si}}_N\nn\\
&&+\tilde{\boldsymbol{\si}}_1\otimes\big[\tilde{\si}_2^{\dg}\tilde{\si}_2,[\tilde{\boldsymbol{\si}}_2,\tilde{\si}_2^{\dg}\tilde{\si}_2]\big]\otimes\tilde{\boldsymbol{\si}}_3\otimes\dots\otimes\tilde{\boldsymbol{\si}}_N+\nn\\
&&\dots+\tilde{\boldsymbol{\si}}_1\otimes\dots\otimes\tilde{\boldsymbol{\si}}_{N-1}\otimes\big[\tilde{\si}_N^{\dg}\tilde{\si}_N,[\tilde{\boldsymbol{\si}}_N,\tilde{\si}_N^{\dg}\tilde{\si}_N]\big]\Big)\nn\\
&&+\G_{\g}\Big(\big(\tilde{\si}_1^{\dg}[\tilde{\boldsymbol{\si}}_1,\tilde{\si}_1]-[\tilde{\boldsymbol{\si}}_1,\tilde{\si}_1^{\dg}]\tilde{\si}_1\big)\otimes\tilde{\boldsymbol{\si}}_2\otimes\dots\otimes\tilde{\boldsymbol{\si}}_N\nn\\
&&+\tilde{\boldsymbol{\si}}_1\otimes\big(\tilde{\si}_2^{\dg}[\tilde{\boldsymbol{\si}}_2,\tilde{\si}_2]-[\tilde{\boldsymbol{\si}}_2,\tilde{\si}_2^{\dg}]\tilde{\si}_2\big)\otimes\tilde{\boldsymbol{\si}}_3\otimes\dots\otimes\tilde{\boldsymbol{\si}}_N+\dots\nn\\
&&+\tilde{\boldsymbol{\si}}_1\otimes\dots\otimes\tilde{\boldsymbol{\si}}_{N-1}\otimes\big(\tilde{\si}_N^{\dg}[\tilde{\boldsymbol{\si}}_N,\tilde{\si}_N]-[\tilde{\boldsymbol{\si}}_N,\tilde{\si}_N^{\dg}]\tilde{\si}_N\big)\Big)\ra,\nn
\eea where $\G_{\g}=\pi\g^{2}/v_g$, $\G_{\lambda}=\pi\lambda^{2}/v_g$.

Next we write down the explicit forms of Eq.~\ref{comm6} for an atomic medium of two atoms. In the presence of losses, the Eqs.~\ref{s8}-\ref{s16} in Sec.~\ref{2atoms} should be replaced by the following equations obtained from Eq.~\ref{comm6}:
\bea
\f{d\mathcal{S}_{1}}{dt}&=&-\big(i \delta \om_1+\tilde{\Gamma}_L+\G_\lambda\big)\mathcal{S}_{1}-4iJ_z\mathcal{S}_{122}\nn\\&&+2iJ_x(2\mathcal{S}_{112}-\mathcal{S}_{2})+2i\Omega_L\mathcal{S}_{11}-i\Omega_L,\label{sd8} \\
\f{d\mathcal{S}_{2}}{dt}&=&-\big(i \delta \om_2+\tilde{\Gamma}_R+\G_\lambda\big)\mathcal{S}_{2}-4iJ_z\mathcal{S}_{112}\nn \\&&+2iJ_x(2\mathcal{S}_{122}-\mathcal{S}_{1}), \label{sd9} \\
\f{d\mathcal{S}_{11}}{dt}&=&-2\tilde{\Gamma}_L\mathcal{S}_{11}-2iJ_x(\mathcal{S}_{12}-\mathcal{S}^*_{12})+i\Omega_L(\mathcal{S}_{1}-\mathcal{S}^*_{1}),\nn\\\label{sd10} \\
\f{d\mathcal{S}_{22}}{dt}&=&-2\tilde{\Gamma}_R\mathcal{S}_{22}+2iJ_x(\mathcal{S}_{12}-\mathcal{S}^*_{12}),\label{sd11}\\
\f{d\mathcal{S}_{3}}{dt}&=&-\big(i(\delta \om_1+\delta \om_2+4J_z)+\tilde{\Gamma}_L+\tilde{\Gamma}_R+2\G_\lambda\big)\mathcal{S}_{3}\nn\\&&+i\Omega_L(2\mathcal{S}_{112}-\mathcal{S}_{2}), \label{sd12}\\
\f{d\mathcal{S}_{12}}{dt}&=&\big(i(\om_1-\om_2)-\tilde{\Gamma}_L-\tilde{\Gamma}_R-2\G_\lambda\big)\mathcal{S}_{12}\nn\\&&+2iJ_x(\mathcal{S}_{22}-\mathcal{S}_{11})+i\Omega_L(\mathcal{S}_{2}-2\mathcal{S}_{112}), \label{sd13}\\
\f{d\mathcal{S}_{122}}{dt}&=&-\big(i(\delta \om_1+4J_z)+\tilde{\Gamma}_L+2\tilde{\Gamma}_R+\G_\lambda\big)\mathcal{S}_{122}\nn\\&&+2iJ_x\mathcal{S}_{112}+i\Omega_L(2\mathcal{S}_{1122}-\mathcal{S}_{22}),\label{sd14} \\
\f{d\mathcal{S}_{112}}{dt}&=&-\big(i(\delta \om_2+4J_z)+2\tilde{\Gamma}_L+\tilde{\Gamma}_R+\G_\lambda\big)\mathcal{S}_{112}\nn\\&&+2iJ_x\mathcal{S}_{122}+i\Omega_L(\mathcal{S}_{3}-\mathcal{S}_{12}), \label{sd15}\\
\f{d\mathcal{S}_{1122}}{dt}&=&-2(\tilde{\Gamma}_L+\tilde{\Gamma}_R)\mathcal{S}_{1122}+i\Omega_L(\mathcal{S}_{122}-\mathcal{S}^*_{122}), \label{sd16}
\eea
where $\tilde{\Gamma}_L=\Gamma_L+\G_\g$ and $\tilde{\Gamma}_R=\Gamma_R+\G_\g$. The rest of the equations are complex conjugates of Eqs.~\ref{sd8},\ref{sd9},\ref{sd12},\ref{sd13},\ref{sd14},\ref{sd15}. We can find the expectation value of the atomic operators from the above equations for some initial condition as we have done in Sec.~\ref{Natoms}.

\begin{figure}
\includegraphics[width=0.99\linewidth]{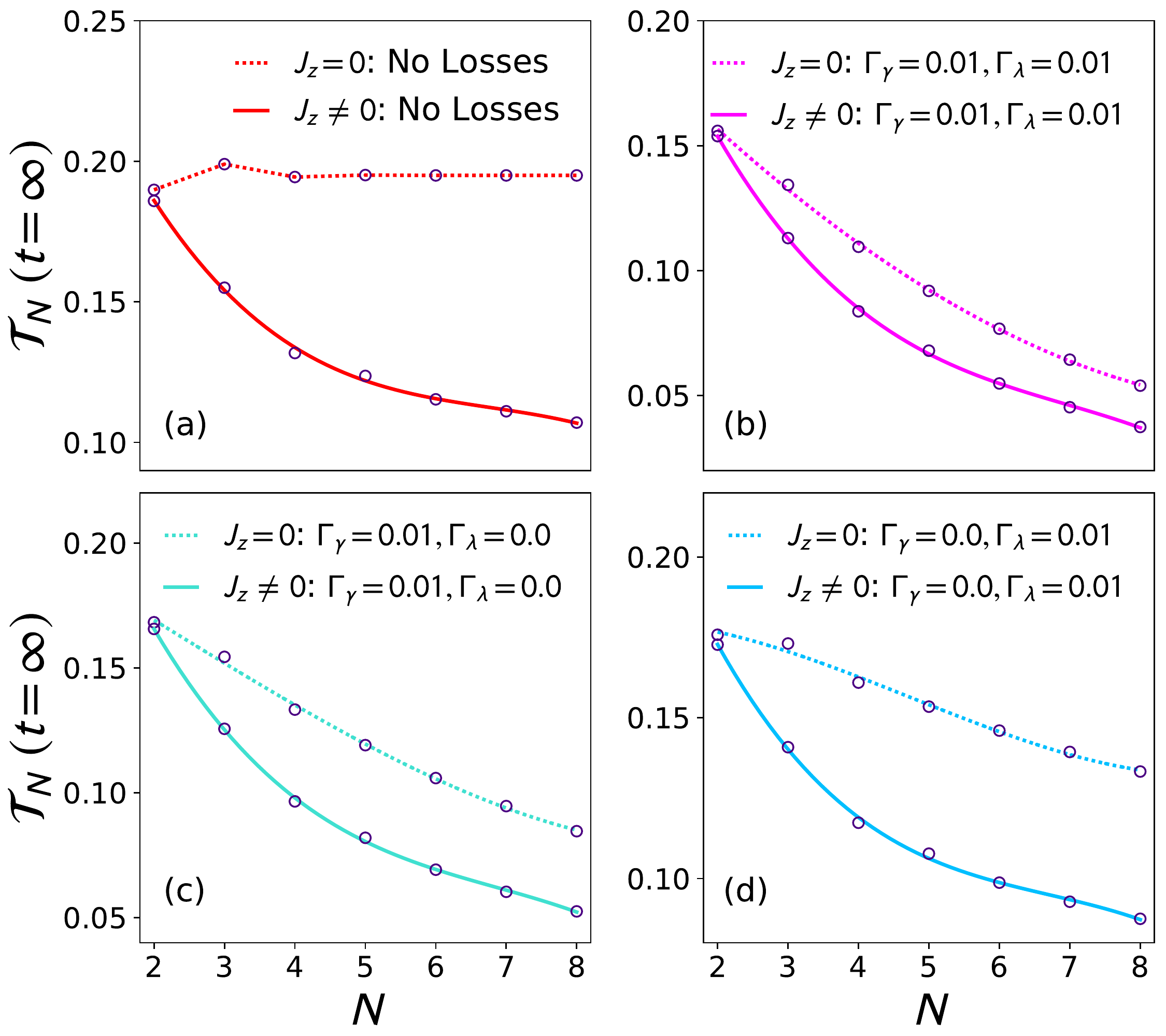}
\caption{Influence of nonradiative decay and pure dephasing on the scaling of nonlinear transmission coefficient $\mathcal{T}_N(t=\infty)$ of a resonant monochromatic laser with the length $N$ of an ordered atomic media modeled as a dissipative XX $(J_z= 0)$ and a dissipative interacting $(J_z \ne 0)$ spin chain. The rates $\Gamma_{\gamma}$ and $\Gamma_{\lambda}$ of the nonradiative decay and the pure dephasing respectively are shown in the plots. The parameters are $I_{\rm in}=0.16\om_a/v_g, \om_p=\om_a, J_x=J_z=0.05\om_a, \Gamma_L=\Gamma_R=0.1\om_a$. The rates $\Gamma_{\lambda},\Gamma_{\gamma}$ are in units of $\om_a$.}
\label{Losses}
\end{figure}

In Fig.~\ref{Losses}, we show how nonradiative decay and pure dephasing affect the scaling of nonlinear transmission coefficient $\mathcal{T}_N(t=\infty)$ of a monochromatic laser with the length $N$ of an ordered atomic medium modeled as a dissipative XX and a dissipative XXZ like interacting spin chain. The nature of nonlinear transmission in the XX spin chain seems to change substantially from ballistic transport in the absence of loss to exponential decay with length in the presence of pure dephasing or nonradiative decay. The fall of nonlinear $\mathcal{T}_N(t=\infty)$ with $N$ in the interacting spin chain at $J_z/J_x=1$ seems to remain algebraic in the presence of pure dephasing and to become exponential in the presence of nonradiative decay. In the presence of both nonradiative decay and pure dephasing, the nonlinear $\mathcal{T}_N(t=\infty)$ falls exponentially with $N$ in the XX and interacting spin chains. However, the above features of $\mathcal{T}_N(t=\infty)$ depend on the rates of these losses and can change at higher rates. 

\section{Long-range interactions}\label{App3}
Here we briefly sketch how to calculate linear and nonlinear light propagation through an atomic medium with long-range interactions between atoms. 
The Hamiltonian of the atomic medium is  
\bea
\f{\mH_{LM}}{\hbar}&=&\sum_{i=1}^{N}\om_i\si_i^\dg\si_i+\sum_{i=1}^{N-1}\sum_{j=i+1}^{N}\big(\f{2J_x}{|i-j|^\al}(\si_i^\dg\si_j+\si_i\si_j^\dg)\nn\\
&&+\f{4J_z}{|i-j|^\be}\si_i^\dg\si_i\si_j^\dg\si_j\big),\label{lr1}
\eea
where the exponents $\alpha$ and $\beta$ control long-rangeness of the couplings. The total Hamiltonian of the atomic medium including the photonic baths and the bath-medium couplings is given by
\bea
\mH_{\rm long}=\mH_{LM}+\mH_{E},
\eea
 and the Hamiltonian $\mH_{E}$ is from Sec.~\ref{Natoms}. We rename $\mH_{LM}$ and $\mH_{\rm long}$, respectively, as $\tilde{\mH}_{LM}$ and $\tilde{\mH}_{\rm long}$ after replacing $\om_i$ in $\mH_{LM}$ by $\delta \om_i=\om_i-\om_p$. 

Single-photon transport in a long-range XX or interacting spin chain can be calculated exactly using the scattering theory discussed in the Appendix~\ref{App1}. For a long-range chain, we have $\boldsymbol{\mathcal{Z}}_{ls}\bold{e}_k=\boldsymbol{\Omega}_{s}$ where $\boldsymbol{\mathcal{Z}}_{ls}$ is a $N \times N$ square matrix whose elements are $[\boldsymbol{\mathcal{Z}}_{ls}]_{jj}=\om_j-v_gk-i\Gamma_{L}\delta_{j,1}-i\Gamma_{R}\delta_{j,N}$ for $j=1,2 \dots N$ and $[\boldsymbol{\mathcal{Z}}_{ls}]_{mn}=2J_x/|m-n|^{\alpha}$ for $m \ne n =1,2\dots N$. $\bold{e}_k$ and $\boldsymbol{\Omega}_{s}$ remain the same as in Eq.~\ref{emat} and Eq.~\ref{omat}. The single-photon transmission and reflection coefficients are then found from Eq.~\ref{stra} and Eq.~\ref{sref} using $\bold{e}_k$ derived from $\bold{e}_k=\boldsymbol{\mathcal{Z}}_{ls}^{-1}\boldsymbol{\Omega}_{s}$. The single-photon transmission is again ballistic in a long-range XX or interacting spin-1/2 chain for any value of $\alpha$ and $\beta$. However, the single-photon transmission coefficient $|t_N(\om_p)|^2$ now oscillates with $N$ at a smaller $\alpha$ even for a resonant light $(\om_p=\om_a)$ and $2J_x=\Gamma_L=\Gamma_R$. The amplitude of this oscillation increases with a decreasing $\alpha$.  

Nonlinear light transmission in a long-range spin chain can be investigated following the QLE method in Sec.~\ref{Natoms}. To find the commutator, 
$[\tilde{\boldsymbol{\si}}^{\otimes N},\tilde{\mH}_{\rm long}]$ for the Heisenberg equations of the atomic operators and hence the expectation value  $-\f{i}{\hbar}\la[\tilde{\boldsymbol{\si}}^{\otimes N},\tilde{\mH}_{\rm long}]\ra$, we note that the commutators involving both the photonic baths and the losses (if included) remain unchanged for a long-range Hamiltonian. Therefore, $\la[\tilde{\boldsymbol{\si}}^{\otimes N},\mH_{E}]\ra$ from Sec.~\ref{Natoms} can also be used here.  We need only compute $\la[\tilde{\boldsymbol{\si}}^{\otimes N},\tilde{\mH}_{LM}]\ra$ to obtain the set of equations  $\f{d \la \tilde{\boldsymbol{\si}}^{\otimes N}\ra}{dt}=-\f{i}{\hbar}\la[\tilde{\boldsymbol{\si}}^{\otimes N},\tilde{\mH}_{\rm long}]\ra$ which can again be cast as,
\bea
\f{d\boldsymbol{\mS}_N(t)}{dt}=\boldsymbol{\mZ}_l\boldsymbol{\mS}_N(t)+\boldsymbol{\Om}_N, \label{EqLR}
\eea
where $\boldsymbol{\mS}_N(t)$ and $\boldsymbol{\Om}_N$ are the same as those in Eq.~\ref{sigma6}, and $\boldsymbol{\mZ}_l$ keeps track of the long-range model.

We can directly compute $\la[\tilde{\boldsymbol{\si}}^{\otimes N},\tilde{\mH}_{LM}]\ra$ by writing $\tilde{\boldsymbol{\si}}^{\otimes N}$ and $\tilde{\mH}_{LM}$ as $2^N\times 2^N$ matrices for $N$ atoms. Here, we write the Hamiltonian $\tilde{\mH}_{LM}$  of the atomic medium as an MPO for an easy numerical implementation. However, there is no straightforward exact method to represent a long-range Hamiltonian by an MPO. Therefore, we here apply an approximation known as the Levenberg-Marquardt nonlinear least squares method \cite{CrosswhitePRB2008}. We can not write the long range Hamiltonian in Eq. \ref{lr1} as a sum of two-site Hamiltonian used in Sec.~\ref{Natoms} due to the separation-dependent coupling between atoms $i$ and $j$. Hence, we make an approximation for the function $1/|i-j|^u$ with $u=\al,\be$ in the MPO representation. The problem then reduces to minimizing,
\bea
&&f(\g_1,\de_1,\g_2,\de_2\dots\g_L,\de_L)\nn\\
&&=\sum_{k=1}^L\sum_{|i-j|=1}^{|i-j|_{max}}\left[\g_k\de_k^{|i-j|-1}-\f{1}{|i-j|^u}\right]^2,
\eea
where $\sum_{k=1}^{L}\g_k\de_k^{|i-j|-1}$ approximates $1/|i-j|^u$ with $u=\al,\be$. One problem with this method is that the dimension of an approximate MPO, $\chi$  increases with the number of coefficients $\gamma_k,\delta_k$. Hence, the dimension of the matrix at each site, $\chi = 3L+2$, is a function of $L$. Let us define the vectors: $\g^{[u]}=[\g_1^{[u]}\:\:\g_2^{[u]}\dots\g_L^{[u]}]$, $\de^{[u]}={\rm diag}[\de_1^{[u]}\mI\:\:\de_2^{[u]}\mI \dots\de_L^{[u]}\mI]$ and $Q=[1\:\:1\dots1]$ of length $L$. For an approximation with $L$ coefficients, we have
\begin{widetext}
\bea
M_l^{[j]}&=& \left[\ba{ccccc}\mI_j& 0 & 0& 0& 0\\Q^T\otimes\si_j&\de^{[\al]}& 0& 0& 0\\Q^T\otimes\si_j^\dg& 0&\de^{[\al]}& 0 & 0\\Q^T\otimes\si_j^\dg\si_j& 0& 0&\de^{[\be]}& 0\\\de\om_j\si_j^\dg\si_j&2J_x\g^{[\al]}\otimes \si_j^\dg& 2J_x\g^{[\al]}\otimes \sigma_j&4J_z\g^{[\be]}\otimes \si_j^\dg\si_j&\mI_j\ea\right], \\
M_l^{[1]}&=&\left[\ba{ccccc}\de\om_1\si_1^\dg\si_1&2J_x\g^{[\al]}\otimes \si_1^\dg& 2J_x\g^{[\al]}\otimes \sigma_1&4J_z\g^{[\be]}\otimes \si_1^\dg\si_1&\mI_1\ea\right],\\
M_l^{[N]}&=&\left[\ba{ccccc}\mI_N&Q\otimes\si_N&Q\otimes\si_N^\dg&Q\otimes\si_N^\dg\si_N&\de\om_N\si_N^\dg\si_N\ea\right]^T,
\eea
\end{widetext}
where $j=2,3\dots N-1$, $\mI_j$ is a $2\times2$ identity matrix and $0$ is a null matrix of required dimensions. The Hamiltonian thus becomes,
\bea
\tilde{\mH}_{LM}=M_l^{[1]}M_l^{[2]}\dots M_l^{[N]}.
\eea
Each $M_l^{[j]}$ represents a matrix in the local Hilbert space of the individual atom $j$ denoted by the superscript. The matrix $\tilde{\mH}_{LM}$ then represents the  Hamiltonian in the complete Hilbert space of the atomic medium. 

\begin{figure}
\includegraphics[width=0.99\linewidth]{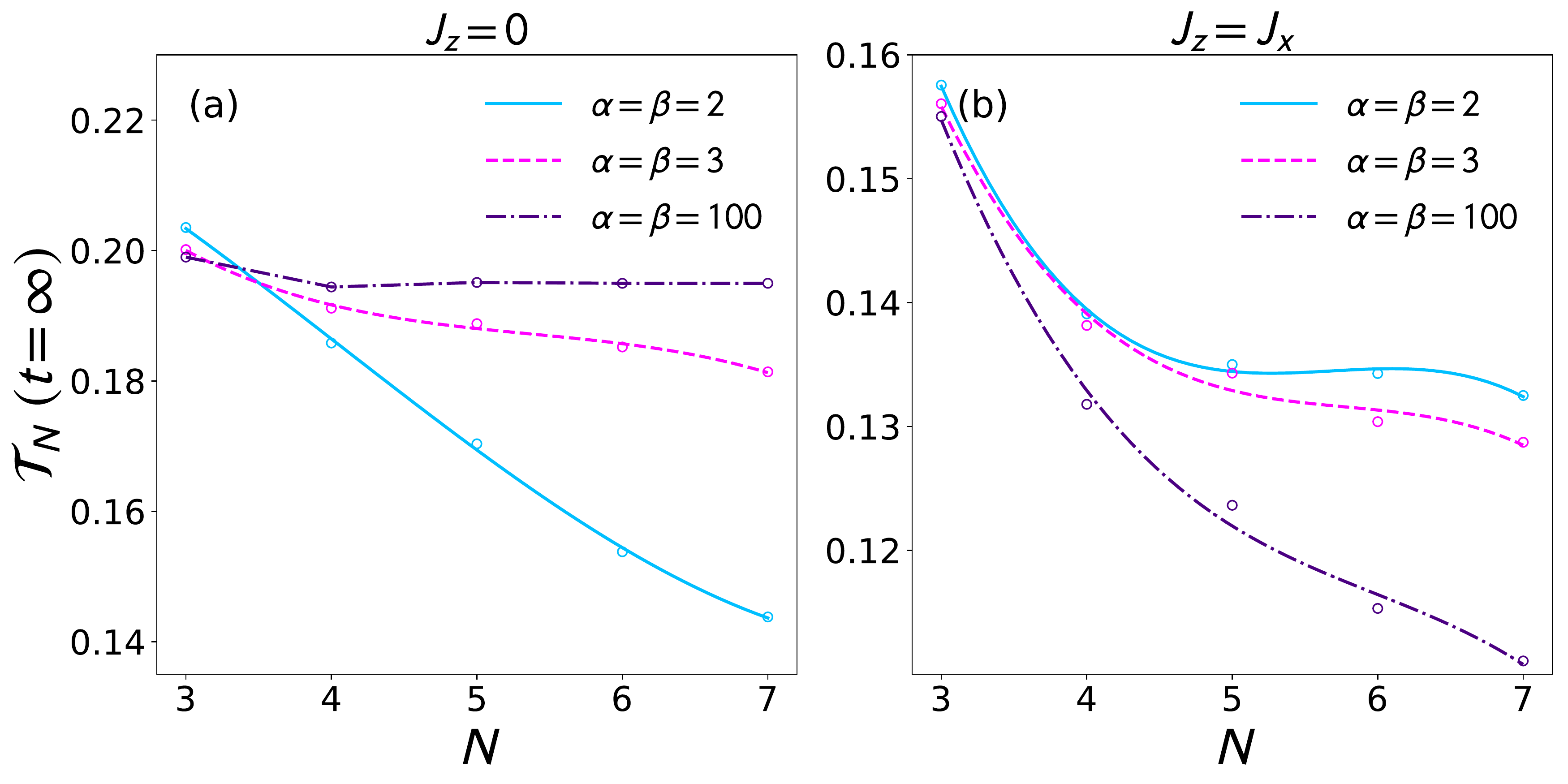}
\caption{Role of long-range coupling in the scaling of nonlinear transmission coefficient $\mathcal{T}_N(t=\infty)$ of a resonant monochromatic laser with the length $N$ of an ordered atomic media modeled as a long-range XX $(J_z= 0)$ and a long-range interacting $(J_z \ne 0)$ spin chain. The exponents $\alpha$ and $\beta$ determine the long-rangeness and they are shown as figure headings. The parameters are $I_{\rm in}=0.16\om_a/v_g, \om_p=\om_a, J_x=J_z=0.05\om_a, \Gamma_L=\Gamma_R=0.1\om_a$.}
\label{LongRange}
\end{figure}

We apply the above approximate method to write and solve the equations in \ref{EqLR}. While we can investigate single-photon transport using the scattering theory for an arbitrary number of atoms, we can calculate nonlinear light transmission only for short chains due to the limitation on the size of $\boldsymbol{\mZ}_l$ in our numerics. In Fig.~\ref{LongRange}, we show how the length-dependence of steady-state transmission $\mathcal{T}_N(t=\infty)$ in a long-range XX and interacting spin chain evolves with $\alpha$ and $\beta$. We find that the nonlinear $\mathcal{T}_N(t=\infty)$ falls with $N$ in the long-range XX chain for small $\alpha$ and $\beta$, and it becomes independent of $N$ for large $\alpha$ and $\beta$ as expected. However, it is difficult to investigate the features of nonlinear light transmission in a long-range model conclusively due to our current numerical constraint.

\bibliography{bibliography1}
\end{document}